\documentclass[3p,times]{elsarticle}
\usepackage{amssymb}

\usepackage{caption}
\usepackage{subfigure}  
\usepackage{amsmath}
\usepackage{bm}
\usepackage{color}
\usepackage[colorlinks=True]{hyperref}
\begin{document}
\begin{frontmatter}
\title{Three-body problem --  from Newton  to supercomputer plus machine learning}
\author[label1,label2,label3]{Shijun Liao \corref{cor1}}
\author[label4,label5]{Xiaoming Li}
\author[label2]{Yu Yang}
\cortext[cor1]{sjliao@sjtu.edu.cn}

\address[label1]{State Key Lab of Ocean Engineering, Shanghai 200240, China}
\address[label2]{Center of Marine Numerical Experiment, School of Naval Architecture, Ocean and Civil Engineering, Shanghai Jiaotong University, Shanghai 200240,  China} 
\address[label3]{School of Physics and Astronomy,  Shanghai Jiaotong University, Shanghai 200240, China}
\address[label4]{School of Mechanics and Construction Engineering, 
Jinan University, Guangzhou 510632, China}
\address[label5]{MOE Key Laboratory of Disaster Forecast and Control in Engineering, Guangzhou 510632, China}

\begin{abstract}
The famous three-body problem can be traced back to Newton in 1687, but quite few families of periodic orbits were found in 300 years thereafter.  In this paper,  we propose an effective approach  and  roadmap  to numerically gain  planar  periodic orbits of three-body systems with arbitrary masses by means of machine learning based on an artificial neural network (ANN) model.  Given any a known periodic orbit as a starting point,  this approach can provide more and more periodic orbits (of the same family name) with  variable masses, while the mass domain having periodic orbits becomes larger and larger,  and the ANN model becomes wiser and wiser.   Finally we have an ANN model trained by means of all obtained periodic orbits of the same family, which provides a convenient way to give accurate enough predictions of periodic orbits with arbitrary masses for  physicists and astronomers.     It suggests that the high-performance computer and artificial intelligence (including machine learning)  should be  the key to gain periodic orbits of the famous three-body problem.  
\end{abstract}	


\end{frontmatter}

\section{Introduction}

How are  the  trajectories  of  three point masses $m_1, m_2$ and $m_3$ that are attracted each other by Newton's gravitational law?   This so-called ``three-body problem''  can be traced back to Newton  \cite{Newton1687} in 1687.  According to Newtion's  second law and gravitational law,  the related governing equations about $N$-body problem read
\begin{equation}
	m_{k}\frac{d^{2 }{\bf r}_{k}}{d t^{2} } = \sum_{j=1,\\ j\neq k}^{N} \frac{G m_{k} m_{j} ({\bf r}_{j}-{\bf r}_{k})}{\left|{\bf r}_{j}-{\bf r}_{k}\right|^{3}},  \hspace{0.5cm} 1\leq k \leq N , \label{geq:Newton}
\end{equation}
where ${\bf r}_{k}$ and $m_{k}$  are the position vector and mass of the $k$th-body,  $t$ denotes the time, respectively, with a given initial position $ {\bf r}_{0,k}$ and velocity ${\bf v}_{0,k}$, i.e. 
\begin{equation}
	{\bf r}_{k}(0) =  {\bf r}_{0,k},  \dot{\bf r}_{k}(0) = {\bf v}_{0,k}, \hspace{0.5cm} 1\leq k \leq N.
	\end {equation}
	Here the dot denotes the derivative with respect to $t$.  Note that  ${\bf r}_{k}$, $m_{k}$ and $t$  are dimensionless using a characteristic length $L$, a characteristic mass $M$ and the Newton's gravitational constant $G$.    If the trajectory of each body at $t = T$ exactly returns its initial status, say, 
	\begin{equation}
		{\bf r}_{k}(T) = {\bf r}_{k}(0) = {\bf r}_{0, k},  \;\;\;  \dot{\bf r}_{k}(T) = \dot{\bf r}_{k}(0) = {\bf v}_{0, k}, \hspace{0.3cm} 1\leq k \leq N,
	\end{equation}  
	one has a periodic solution of the $N$-body problem.   The famous three-body problem   \cite{Poincare1890, Broucke1975a, Broucke1975b, Hadjidemetriou1975a, Henon1976, More1993, Montgomery1998, Li2017, Li2018, Sun2018kepler, Kwiecinski2018IJBC, Stone2019, Li2019NA, Lin2019PRD, Tanikawa2019CMDA, Abouelmagd2020NA, Li2020} corresponds to $N$ = 3.  


\section*{1.1 Times of Newton, Euler, Lagrange  and Poincar\'{e}} 
	
	For the two-body problem, corresponding to  $N$ = 2,  Newton gave a closed-form periodic solution.   However,  for a three-body problem ( $N$ =  3 ), it becomes extremely difficult to find periodic orbits:   no periodic orbits had been found until  Euler reported one in 1740  and Lagrange published one in 1772.  However,  according to Montgomery's  topological method  \cite{Montgomery1998} of classifying  periodic orbits of three-body systems, they belong to the {\em same} family, namely the ``Euler - Lagrange family''.  Thereafter,  {\em no} new periodic orbits were reported in about two centuries.     
	
	Why is it so difficult to find periodic orbits of three-body systems?   The mysterious reasons were  revealed by Poincar\'{e}  \cite{Poincare1890}  in 1890, who proved that, unlike the two-body problem that is integrable and thus its solutions is completely understood,  the three-body problem is {\em not} integrable.   This well explains why {\em only} the Euler - Lagrange family (in closed form) were found in more than two hundred years, since closed-form orbits do {\em not} exist at all in general.   It implies that one generally  {\em had} to use numerical algorithms to solve three-body problem, but unfortunately the electronic computer was even {\em not} invented in the times of  Poincar\'{e}  \cite{Poincare1890}.     In addition,  Poincar\'{e}  \cite{Poincare1890} also found that trajectories of  three-body system  are  generally rather sensitive to initial conditions (i.e. the butterfly-effect), which leaded to the foundation of a  new field of modern science, i.e.  chaotic dynamics.    Nowadays,  it is well-known that, due to the famous butterfly-effect, i.e.  a hurricane in North America might be created by a flapping of wings of a distant butterfly in South America several weeks earlier,  it is very difficult to gain reliable trajectories of chaotic systems, especially in a long interval of time.   All of these  explain why periodic orbits of three-body problem are so difficult to obtain and why it becomes one of the oldest open question in science.    
	
	In most cases, trajectories of three-body system are chaotic, i.e. non-periodic, as discovered by Poincar\'{e}   \cite{Poincare1890}  in 1890 and confirmed again by Lorenz  \cite{Lorenz1963} in 1963 and Stone \&  Leigh  \cite{Stone2019} in 2019.   However, in some special cases, there indeed exist periodic orbits,  i.e.  the three bodies  exactly  return to  their initial positions and initial velocities after a period $T$.    The periodic orbits of the three-body problems are very important, since they are ``the only opening through which we can try to penetrate in a place which, up to now, was supposed to be inaccessible'', as pointed out by Poincar\'{e}  \cite{Poincare1890}.    However, the question is: how can we find them {\em effectively}?

	\section*{1.2 Times of supercomputer}
	
	The  excellent work of  Poincar\'{e}  \cite{Poincare1890} made a historical turning point of searching for periodic orbits of three-body system.   The non-existence of the uniform first integral of triple system reveals the impossibility of finding closed-form analytic solutions of three-body problems in general cases: it clearly indicates that we mostly {\em had to} (i.e. must) use numerical algorithms to solve this problem.   This was indeed a revolutionary contribution of Poincar\'{e}   \cite{Poincare1890} with great foresightness at that times when there was even  {\em no} electronic computers at all:  about a half century later, Turing   \cite{Turing1936, Turing1950}  published  his  epoch-making  papers  that  became  the foundation of modern computer and artificial intelligence.    Thanks  to  Von Neumann   \cite{VonNeumann1958}, who proposed the so-called Von Neumann - Machine for modern computer, also to Jack S. Kilby,  who won the Nobel prize in Physics in 2000 for his taking part in the invention of the integrated circuit,  and to many scientists, mathematicians, engineers and so on whose names we have no  space  to mention here,  the performance of  supercomputer becomes better and better, together with more and more advanced numerical algorithms.   This provides us a  strong  support   of  hardware and software  for  discovering  new  periodic orbits  of  three-body  problem  now.  
	
	With the ceaseless  progress of electronic computers, more and more researchers followed  the  way,  which Poincar\'{e}     \cite{Poincare1890} suggested,   to  numerically search for periodic orbits of three-body problem.    According to Montgomery's  topological method   \cite{Montgomery1998} of classifying  periodic orbits of three-body systems,  only three families of periodic orbits were  found in 300 years after Newton, i.e. 
	(1) the so-called  ``Euler - Lagrange family'' found by Euler in 1740 and Lagrange in 1772;
	(2) the so-called  ``BHH  family'' {\em numerically} found by Broucke   \cite{Broucke1975a, Broucke1975b} in 1975, Hadjidemetriou  \cite{Hadjidemetriou1975a} in 1975 and H\'{e}non  \cite{Henon1976} in 1976,  respectively;   
	(3) the so-called  ``figure-eight family''  for three equal masses {\em numerically} found by Moore   \cite{More1993} in 1993, until 2013 when \v{S}uvakov and Dmitra\v{s}inovi\'{c}   \cite{Suvakov2013} {\em numerically} gained 11 new families of periodic orbits of triple system with three equal masses.  All of these suggest  that  finding periodic orbits of three-body problem by means of numerical methods should be a correct way.  
	
Note that, among the three families of periodic orbits mentioned above,  the BHH family and figure-8 family were found by numerical methods respectively in 1975s and 1993,  when the performance of computer might be not good enough.   However, in 2010s   we  had supercomputer with peak performance about 1,000 petaflops, i.e.  several billion billion fundamental calculations per second, together with lots of powerful numerical algorithms.   What  prevents  us  from  {\em effectively} finding  thousands of  new families of periodic orbits of three-body problem though we have so powerful supercomputer ?    
	
	The key of finding periodic orbits is to gain reliable computer-generated  trajectories of three-body system under arbitrary initial conditions in a {\em long enough} interval of time.   However, as discovered by  Poincar\'{e}    \cite{Poincare1890} and  rediscovered by Lorenz  \cite{Lorenz1963},  computer-generated trajectories of chaotic systems are sensitive to initial conditions, i.e. the famous butterfly-effect.  In other words,  a tiny difference on initial conditions might lead to a huge deviation of computer-generated simulation after a long  time.
	In addition,  Lorenz  \cite{Lorenz1989, Lorenz2006} further found that computer-generated trajectories of chaotic systems  are also sensitive to algorithms: different numerical algorithms might  give distinctly  different computer-generated   trajectories of chaotic systems after a long  time.  This is indeed a great obstacle!   This kind of sensitive dependence on numerical algorithm (SDNA) for a chaotic system had been also observed and reported by many researchers  \cite{jianping2000, jianping2001, Teixeira2005},  which  however  unavoidably  leaded  to  some  intense  arguments  \cite{Yao2008}: some researchers even suggested that ``all chaotic responses should be simply numerical noises'' and might ``have nothing to do with differential equations''.   Besides, it is currently reported by \cite{Chandramoorthy2021JCP} that ``shadowing solutions can be almost surely nonphysical'', which ``invalidates the argument that small perturbations in a chaotic system can only have a small impact on its statistical behavior''.  
	Thus, by means of traditional numerical algorithms (mostly in double precision),  it is rather  difficult to gain reliable/convergent computer-generated trajectories of three-body system under arbitrary initial conditions in a  long {\em enough} interval of time.  
	
	To overcome this obstacle,  Liao  \cite{Liao2009} proposed the so-called  ``clean numerical simulation'' (CNS) for chaotic systems.   Unlike other traditional numerical algorithms, which mostly use double precision,  the CNS  can greatly reduce {\em not only} truncation error {\em but also} the round-off error to keep the total ``numerical noises'' in such a required tiny level that a reliable (or convergent) computer-generated simulation can be obtained in a long enough interval of time.   In the frame of the CNS, the truncation error is decreased by means of numerical algorithms at high enough order in time and space,  and the round-off error is reduced by using multiple-precision with many enough digits for all parameters and variables.   The CNS has been successfully applied to many chaotic systems, such as Lorenz equation, chaotic  three-body systems,  some spatio-temporal  chaotic systems  and so on  \cite{Liao2013-3b, Liao2014-SciChina, Liao2015-IJBC, Hu2020,Liao2022AAMM}.   As reported by Hu and Liao  \cite{Hu2020},  the use of double precision  might lead to huge deviations  of  computer-generated simulations of spatio-temporal  chaos  {\em even}  in statistics, not only quantitatively but also qualitatively, particularly in a long interval of time.   This indicates that we must be very careful to numerically simulate chaotic systems.    Fortunately, the  CNS can provide us a guaranteed tool to gain reliable/convergent trajectories of  chaotic systems (such as three-body systems with arbitrary initial conditions) during a long enough time.  
	
	Using the CNS as an integrator of the governing equations and combining the  grid  search  method  and the Newton-Raphson method  \cite{Farantos1995, Lara2002,  Abad2011},   Li and Liao  \cite{Li2017} in 2017 successfully found 695 families of periodic planar collisionless  orbits of three-body systems with {\em three} equal masses and zero angular momentum,  including the figure-eight family, the 11 families found by \v{S}uvakov and Dmitra\v{s}inovi\'{c} in 2013, and besides more than 600 new families that have never been reported.   Similarly,  Li, Jing and Liao  \cite{Li2018}  further found  1349 new families of  periodic planar collisionless  orbits of the three-body system with only {\em two} equal masses.    In 2020,  starting from a known periodic orbit with three equal masses and using the CNS to integrate the governing equations,  Li {\em et al.}   \cite{Li2020} successfully obtained 135445 new periodic orbits with arbitrarily {\em unequal} masses by means of combining  the numerical continuation method  \cite{Allgower2003} and the Newton-Raphson method  \cite{Farantos1995, Lara2002,  Abad2011}, including 13315 stable ones.   Therefore, in only four years,  using high-performance computer and our new strategy based on the CNS,  we successfully increased the  family number of the known periodic orbits of three-body systems by nearly {\em four} orders of magnitude!    This strongly indicates that our numerical strategy is correct, powerful and rather {\em effective} for finding new periodic orbits of  three-body systems.   It should be emphasized that this great progress is mainly due to the use of the new numerical strategy based on the CNS, since the performance of supercomputer is good enough  for three-body systems  even at the beginning of the 21st century.

Triple stars systems are key objects in astronomy. 
All observed periodic  triple stars belong to the BHH family   \cite{Jankovic2016}. So it is an open problem whether there exist other families of periodic triple stars in our universe. In this paper, we present a new method to obtain periodic orbits of the three-body problem with different masses. This will enrich our understanding about the triple system. 
	\section{An approach based on machine learning}
	
	However,  it is time-consuming to use the numerical continuation method  \cite{Allgower2003} to find the 135445 periodic orbits (with {\em unequal} masses) reported by Li, Li and Liao  \cite{Li2020}. 

The  numerical continuation method was applied to obtain solutions of differential equations
$
\dot{\bm{u}} = F(\bm{u},\lambda),
$
where $\lambda$ is a  physical parameter, also called ``natural parameter''. Assume $\bm{u}_0$ is a solution of the system at a natural parameter  $\lambda = \lambda_0$.  Using this solution $\bm{u}_0$ at  $\lambda = \lambda_0$ as an initial guess, a new solution $\bm{u}'$  can be obtained at a new natural parameter $\lambda = \lambda_0 + \Delta \lambda$ through the Newton-Raphson method  \cite{Farantos1995, Lara2002} and the clean numerical simulation (CNS)  \cite{Liao2009,  Liao2014-SciChina,  Hu2020},  if the increment $\Delta \lambda$ is small enough to make sure iterations convergence.
	 Besides, these periodic orbits are in essence {\em discrete}, say,  only for some {\em specific} values of $m_1$ and $m_2$ in an irregular domain (in case of $m_3 = 1$ since we use the mass of the 3rd body as a characteristic mass $M$, without loss of generality).   Can we gain a periodic orbit more {\em efficiently} for {\em arbitrary} values of masses $m_{1}$ and $m_{2}$?   Thanks to the times of machine learning, the answer is positive and rather attractive, as described below.          
	For example,  let us use a known BHH orbit  as a starting point to illustrate this.   
	Fifty-seven  satellite periodic orbits of the BHH family of three-body systems with three equal masses were found  \cite{Jankovic2016} in 2016,  and this number was extended to ninety-nine  \cite{Jankovic2020} in 2020.  The initial configuration of the BHH satellite periodic orbits with zero angular momentum is described by 
	\[  \bm{r}_1(0)=(x_1,0),  \bm{r}_2(0)=(1,0),   \bm{r}_3(0)=(0,0), \] 
	and 
	\[  \dot{\bm{r}}_1(0)=(0,v_1),  \dot{\bm{r}}_2(0)=(0,v_2), \]
	\[ \dot{\bm{r}}_3(0)=(0, -(m_1v_1+m_2v_2)/m_3),\]
	where $\bm{r}_i$, $\dot{\bm{r}}_i$ and $m_i$ is the position vector, velocity vector and mass of the $i$-th body, respectively.    Thus, for given $m_{1}, m_{2}$ and $m_{3}$ (we assume $m_3 = 1$ thereafter), we should determine four  unknown physical variables $x_{1},v_{1}, v_{2}$ and the period $T$.  Note that these orbits are periodic  in a rotating frame of reference, say,  the frame of coordinates rotates an angle  $\theta_{T}$ in the corresponding period $T$.  

\captionsetup{belowskip=-0.3pt}
	
	\begin{figure}[t]
		\centering
		\includegraphics[width=8.7cm]{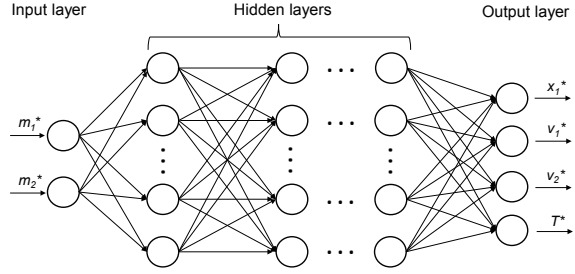}		
		\caption{The artificial neural network (ANN) model, where the input $(m_1^*, m_2^*)$ and output $(x_1^*, v_1^*, v_2^*,  T^*)$ are the normalized data of $(m_1, m_2)$ and  $(x_1, v_1, v_2, T)$, respectively.}
		\label{fig-NN}
	
	\end{figure}

 Without loss of generality, let us consider such a known BHH periodic orbit with the initial condition 	 
$x_1$=-1.325626981682458, $v_1$=-0.8933877752879044, $v_2$=-0.2885702941263346, 
the period $T=9.199307755830397$ and the rotation angel $\theta_{T} = 0.383160887655628$ 
 of the coordinate frame, where $m_1=m_2=m_3=1$.   Using this periodic orbit as a starting point, we first follow Li, Li and Liao  \cite{Li2020} to obtain only 36 periodic orbits for different masses in a small domain  $m_1\in [0.95, 1.00], m_2 \in [1.00, 1.05]$ (marked by $S_1$, with the mass increment $\Delta m_1 = \Delta m_2 = 0.01$) by means of the numerical continuation method  \cite{Allgower2003} and the Newton-Raphson method  \cite{Farantos1995, Lara2002,  Abad2011}.  
Firstly, using the periodic orbit with equal masses,  the periodic orbits with different masses $m_1$ and $m_2=m_3=1$ can be obtained by means of the continuation method and the Newton-Raphson method. Secondly, using these periodic orbits with different masses $m_1$ and $m_2=m_3=1$ as starting points, we obtain periodic orbits with different $m_2$ by means of the continuation method and the Newton-Raphson method. Finally, we can obtain periodic orbits with the mass domain  $m_1\in [0.95, 1.00], m_2 \in [1.00, 1.05]$ and $m_3=1$.  
For details, please refer to Li, Li and Liao  \cite{Li2020}.  The return distance (deviation) of these periodic orbits is defined  by 
	\[  \delta_{T}=\sqrt{\sum_{i=1}^3\Big( \left|\bm{r}_i(T)-\bm{r}_i(0)\right|^2+\left|\dot{\bm{r}}_i(T)-\dot{\bm{r}}_i(0)\right|^2\Big)} \]
	in the rotating frame of reference, where $T$ is the period.   Note that all of the 36 periodic orbits satisfy the criteria $\delta_{T}< 10^{-10}$.  Besides, they all have the same family name, which is defined by the so-called ``free group element''  according to Montgomery's  topological method  \cite{Montgomery1998}.

\begin{figure}[tb]
	\centering
	\includegraphics[width=8cm]{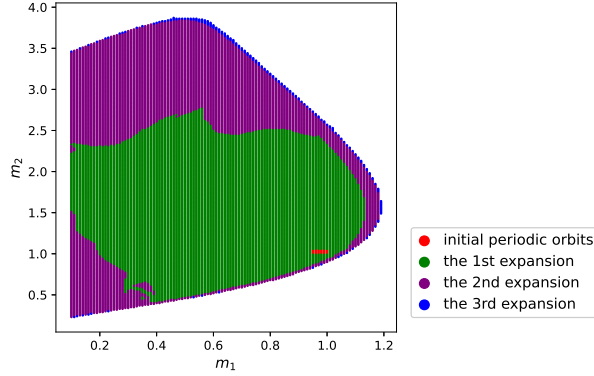}
	\caption{The periodic orbits found by expansion on the mass region. Red dot ($S_1$): initial periodic orbits; green dot ($S_2$): the first extrapolation/expansion  ; purple dot ($S_3$): the second extrapolation/expansion; blue dot ($S_4$): the third extrapolation/expansion. }
	\label{expand-arc}
\end{figure}

\begin{figure}[t]
	\centering
	\subfigure[]{\includegraphics[width=6cm]{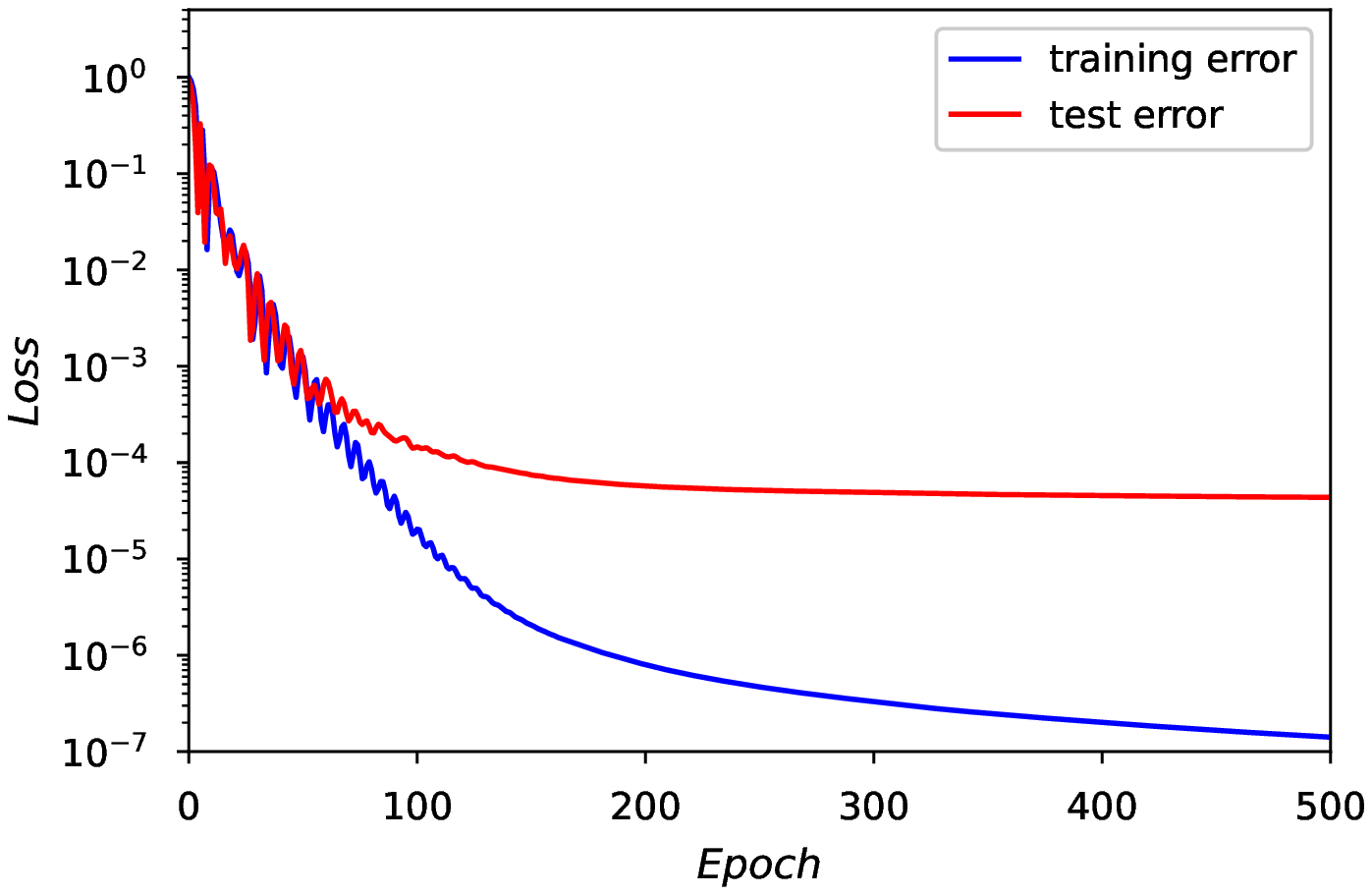}}
	\subfigure[]{\includegraphics[width=6cm]{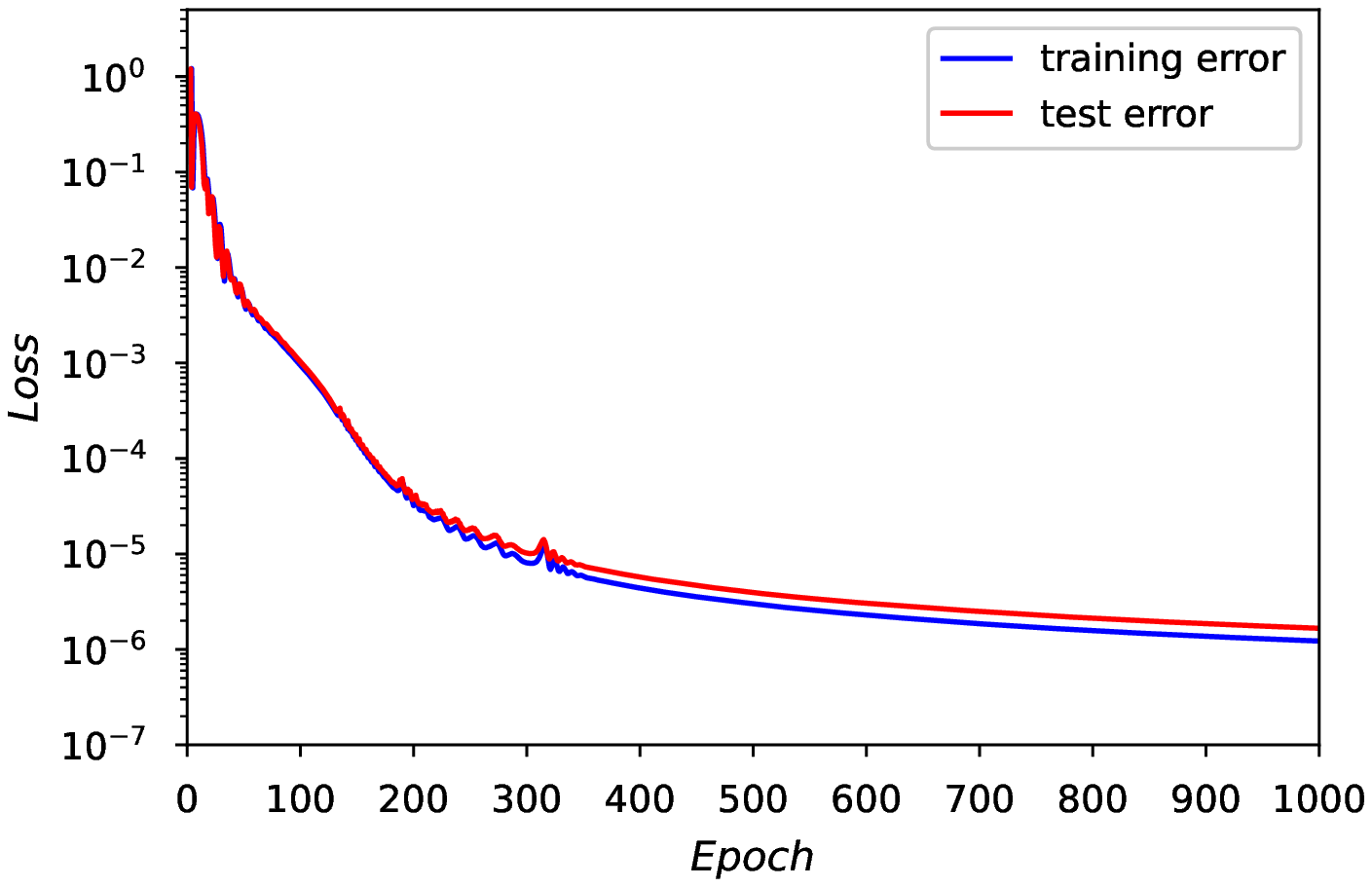}}
	\vspace{0.1cm}
	\caption{(a) Loss function of the neural network for the first training with 36 periodic orbits. (b) Loss function of the neural network for the second training with 17457 periodic orbits. Blue line: training error. Red line: test error.}
	\label{lossplot}
\end{figure}		
	Artificial neural network   \cite{hassoun1995fundamentals, gevrey2003review, andrews1995survey, abiodun2018state} is a machine learning technique evolved from the idea of simulating the human brain, which can be used to perform statistical modelling. Compared with traditional regression approaches, the main advantages of ANN are as follows. First, the ANN does not require information about the complex nature of the underlying process to be explicitly expressed in mathematical form   \cite{sudheer2002data}.  Besides, the ANN  has capability of modelling more complex nonlinear relationships   \cite{livingstone2008artificial}. Therefore, generally speaking, the ANN is applicable across a wider range of problems than the traditional regression approaches. Especially, the ANN can easily deal with  classification problems with complicated boundary  \cite{bala2017classification}. Thus, we use the ANN in the following parts to model the relationship between the parameters of three-body problem and to classify the types of orbits and their stability.
	
	In the next step, we use the ANN model to gain a relation between the input vector  ($m_1$, $m_2$) and  the output vector ($x_1$, $v_1$, $v_2$, $T$) of the periodic orbits.   The ANN  model we used here consists of multiple fully connected layers, say, one input layer, six hidden layers and one output layer.  The number of neurons for the input layer, the hidden layers and the output layer is 2, 1024 and 4, respectively, as shown in Figure~\ref{fig-NN}.   We use  the  optimization  algorithm  AMSGrad  \cite{reddi2019convergence}  as an optimizer  to minimize the mean square error for training the ANN model.   At beginning,   we use the results of the 36 known periodic orbits gained by means of  the numerical continuation method  \cite{Allgower2003} and the Newton-Raphson method  \cite{Farantos1995, Lara2002,  Abad2011} as the training set to train the ANN model. 
	The trained ANN  model provides us a kind of  relationship (expressed by ${\cal F}_1$) between $(m_1, m_2)$ and $(x_1, v_1, v_2, T)$, which can be further used to predict  the initial conditions $x_1$, $v_1$, $v_2$ and the period $T$ of candidates (i.e. possible periodic orbits)  for various masses ($m_1, m_2$) {\em outside} of the original small  domain  $S_1 = \left\{ (m_1, m_2): m_1 \in [0.95, 1.00], m_2 \in [1.00, 1.05]\right\}$.   
	Using the neural network's predictions as the initial guesses for the Newton-Raphson method, we first obtain the periodic orbits with various $m_2$ when $m_1=1$, and with various $m_1$ when $m_2=1$ in order to give a reference of the mass region of possible candidates. Then, we set the candidates of periodic orbits within the mass region $(m_1, m_2)\in[0.1,1.2),[0.4,2.8))$ with $\Delta m=0.01$ and exclude the mass region where the periodic orbits have been found. Finally, we find the 17421 new periodic orbits (66.1$\%$) within this region of 26364 candidates with $\delta_{T}< 10^{-10}$ for various $(m_1, m_2)$ {\em outside} of $S_1$, which are marked in green in Figure~\ref{expand-arc}  and expressed by $S_2$. In the same way,  we further use the results of {\em all} (i.e. 36 + 17421  = 17457)  known  periodic orbits as a training set to train our ANN model so as to gain a better relationship ${\cal F}_2$ between ($m_1$, $m_2$) and  ($x_1$, $v_1$, $v_2$, $T$) for extrapolation/expansion {\em outside} of the mass domain. We set the candidates of periodic orbits within the mass domain $(m_1, m_2)\in ([0.1,1.2),[0.2,3.9))$ with $\Delta m=0.01$ and exclude the mass region where the periodic orbits have been found. Finally, we find the new 11473 periodic orbits marked in purple in Figure~\ref{expand-arc} and expressed by $S_3$ about 49.4$\%$ of all the 23243 candidates.  Similarly, we use the results of  {\em all} (i.e. 36 + 17421  + 11473 = 28930 ) known periodic orbits as the training set to further train our ANN model so as to give a relationship ${\cal F}_3$ between $(m_1, m_2)$ and $(x_1, v_1, v_2, T)$. The chosen mass region $(m_1, m_2)$ of possible candidates is just beyond the boundary of previous obtained periodic orbits, where the horizontal or vertical distance from the $(m_1, m_2)$ to the boundary is less than $7\Delta m (\Delta m=0.01,m_1>0.1,m_2>0.1)$. Then,  we find the 220 new periodic orbits as marked in blue in Figure~\ref{expand-arc} and expressed by $S_4$, only about 7.1$\%$ out of all the 3104 candidates.  
    This indicates that we come to the boundary of  the mass domain $(m_{1},m_{2})$ where the periodic orbits exit.  In other words, there are no periodic orbits beyond that boundary.   Thus, we find the nearly largest mass domain $S^* = S_1 \cup S_2 \cup S_3 \cup S_4$ for the existence of periodic orbits with  some similar properties.

     Note that, for each training, we have examined that there is no overfitting phenomenon before we use the neural network for further prediction. As we randomly divide the whole dataset into the training set (90$\%$) and test set (10$\%$), and find that there is no distinct difference between the errors of these two sets. Then, we train the ANN with all the data including the training and test set to make the best use of the known data for further prediction. For example, for the first training, we randomly divide the dataset with 36 examples into training set  (90$\%$) and test set (10$\%$). The loss function of training is shown in Fig.~\ref{lossplot}(a). The loss is defined as the mean squared error of the standardized data.  The mean relative errors in the training set and test set are about $8.3\times10^{-7}$ and $5.1\times 10^{-5}$, respectively.   
    We find that as the training epoch increases, the error in the test set first decreases and then converges. It is not the phenomenon of overfitting where the loss function of test set first decreases and then increases. As the number of examples is limited, the test error is small while not as small as the training error. Then, we combine the test set into the training set to make use of all the data and use it for further extrapolation prediction.   Note that, unlike the traditional use of the ANN,  we apply the ANN model here to make {\em extrapolation} predictions of periodic orbits.  As for the prediction error of {\em extrapolation} for $(x_1, v_1, v_2, T)$, we calculate the mean relative error by comparing the ANN's predictions with the ``exact'' results which are obtained by modifying the ANN's predictions via the Newton-Raphson method.  The mean relative error of the ANN's extrapolation is about 0.05 for the first expansion of 17421 new periodic orbits.   
    For the second training with 17457 (=17421+36) periodic orbits,  we randomly divide the whole dataset into the training set (90\%) and test set (10\%). The loss function is shown in Fig.~\ref{lossplot}(b).  The mean relative errors in the training set and test set are about $2.2\times10^{-5}$ and $2.3\times 10^{-5}$, respectively.   Thus, as the number of data is sufficient, the difference between the errors of training set and test set is very close. Then,  we train all the 17457 examples for further prediction.  The same approach is repeated in the similar way until no periodic orbits can be found by the ANN's extrapolation prediction.   The mean relative error of the ANN's extrapolation prediction for the second and third expansion are about 0.009 and 0.00007, respectively.
	  As shown in Figure~\ref{expand-arc}, starting from the 36 periodic orbits in the original small domain $(m_1, m_2)\in S_1$ (marked in red), we totally gain the 29150 periodic orbits in the mass domain $(m_1, m_2) \in S^* = S_1 \cup S_2 \cup S_3 \cup S_4$ by the three times extrapolations/expansions. Finally, training our ANN model by using {\em all}   (i.e. 29150) these known results as a training set,  we obtain a  relationship ${\cal F}^*$ between $(m_1, m_2)\in S^*$  and  ($x_1$, $v_1$, $v_2$, $T$), which can give the ANN's predictions of periodic orbits  for arbitrary  values of  $(m_1, m_2)\in S^*$  in the accuracy level of $10^{-4}$ for the return distance (deviation) $\delta_T$.   All of these ANN's predictions of periodic orbits are accurate enough for normal use, although their accuracy can be further modified to arbitrary level by means of the  the Newton-Raphson method via the CNS and multiple-precision (MP), as mentioned below.  Note that the mass domain $(m_1, m_2)$ with the periodic orbits becomes larger and larger, i.e.  from $S_1$ to $S_1 \cup S_2$ to $S_1 \cup S_2  \cup S_3$ to $S^*=  S_1 \cup S_2 \cup S_3 \cup S_4$,  and the relationship between $(m_1, m_2)$ and  ($x_1$, $v_1$, $v_2$, $T$) becomes better and better, from ${\cal F}_1$  to ${\cal F}_2$ to ${\cal F}_3$ to ${\cal F}^*$,  implying that our ANN model becomes wiser and wiser!    

	Obviously, the smaller the return distance (deviation) $\delta_{T}$ is, the more accurate the periodic orbit given by the numerical strategy.   Note that $\delta_{T} = 0$ exactly  corresponds to a closed-form periodic orbit.  But unfortunately,  except the Euler - Lagrange family, nearly all known periodic orbits of three-body systems were gained by numerical methods, and  this fact  is  consistent with  Poincar\'{e}'s  famous proof of the non-existence of the uniform first integral of three-body systems  \cite{Poincare1890}.  It is found that, since the CNS and the MP (multiple precision) are used in our numerical strategy,  the return distance (deviation) $\delta_{T}$ of these periodic orbits can be reduced to {\em any} given value, say, the initial conditions $(x_{1}, v_{1}, v_{2})$  and the period $T$ of these relatively periodic orbits can be at {\em arbitrary} level of precision, i.e.  as accurate as one would like, as long as the performance of supercomputer is good enough.  

	In the present work, we use the ANN model as it is capable of modelling the complex nonlinear relationship,  compared with the traditional approaches such as the linear model or two-dimensional regression. In the first training with 32 data, as the data is limited and the relationship is simple, the traditional approaches indeed have smaller mean relative errors in the test set,  as shown in Table \ref{compare1}.  However, in the second training with 17457 data when the data is sufficient and the relationship between the outputs and the inputs becomes more complex, the ANN model has better performance with smaller error in the test set than the linear model and  the two-dimensional regression, as shown in Table \ref{compare2}.  It reflects that one of the advantages of ANN is its ability to model more complex relationship.  Thus, the ANN can be applicable across a wider range of problems. 
	
		\begin{table*}[t]
		\tabcolsep 0pt \caption{ Mean relative errors of different models for the first training with 36 data.} \small \label{compare1} 
		\begin{center}
			\footnotesize
			\def\temptablewidth{1\textwidth}
			{\rule{\temptablewidth}{1pt}}
			\begin{tabular*}{\temptablewidth}{@{\extracolsep{\fill}}lccc}
				Model & Training set (32 data)  & Test set (4 data)  \\
				\hline
				Linear model	&	3.09$\times 10^{-5}$	&	3.40$\times 10^{-5}$	        	\\
				Two-dimensional interpolation	&	1.95$\times10^{-7}	$&	2.51 $\times 10^{-7}$	       	  	\\
				ANN	&	8.32 $\times 10^{-7}$	&	5.11 $\times 10^{-5}	$       	  	\\
	
			\end{tabular*}
			{\rule{\temptablewidth}{1pt}}
		\end{center}
	\end{table*}
	
	\begin{table*}[t]
		\tabcolsep 0pt \caption{Mean relative errors of different models for the second training with 17457 data.} \small \label{compare2} 
		\begin{center}
			\footnotesize
			\def\temptablewidth{1\textwidth}
			{\rule{\temptablewidth}{1pt}}
			\begin{tabular*}{\temptablewidth}{@{\extracolsep{\fill}}lccc}
				Model & Training set (15711 data)  & Test set (1746 data)  \\
				\hline
				Linear model	&	$1.54\times 10^{-2}$	&	$1.54\times 10^{-2}	$        	\\
				Two-dimensional interpolation	&	$2.49\times 10^{-3}$	&	$2.49\times 10^{-3} $     	  	\\
				ANN	&	$2.18\times 10^{-5}$	&	$2.26\times 10^{-5}$      	  	\\
	
			\end{tabular*}
			{\rule{\temptablewidth}{1pt}}
		\end{center}
	\end{table*}

	\begin{figure*}[t]
		\centering
		\includegraphics[width=4.2cm]{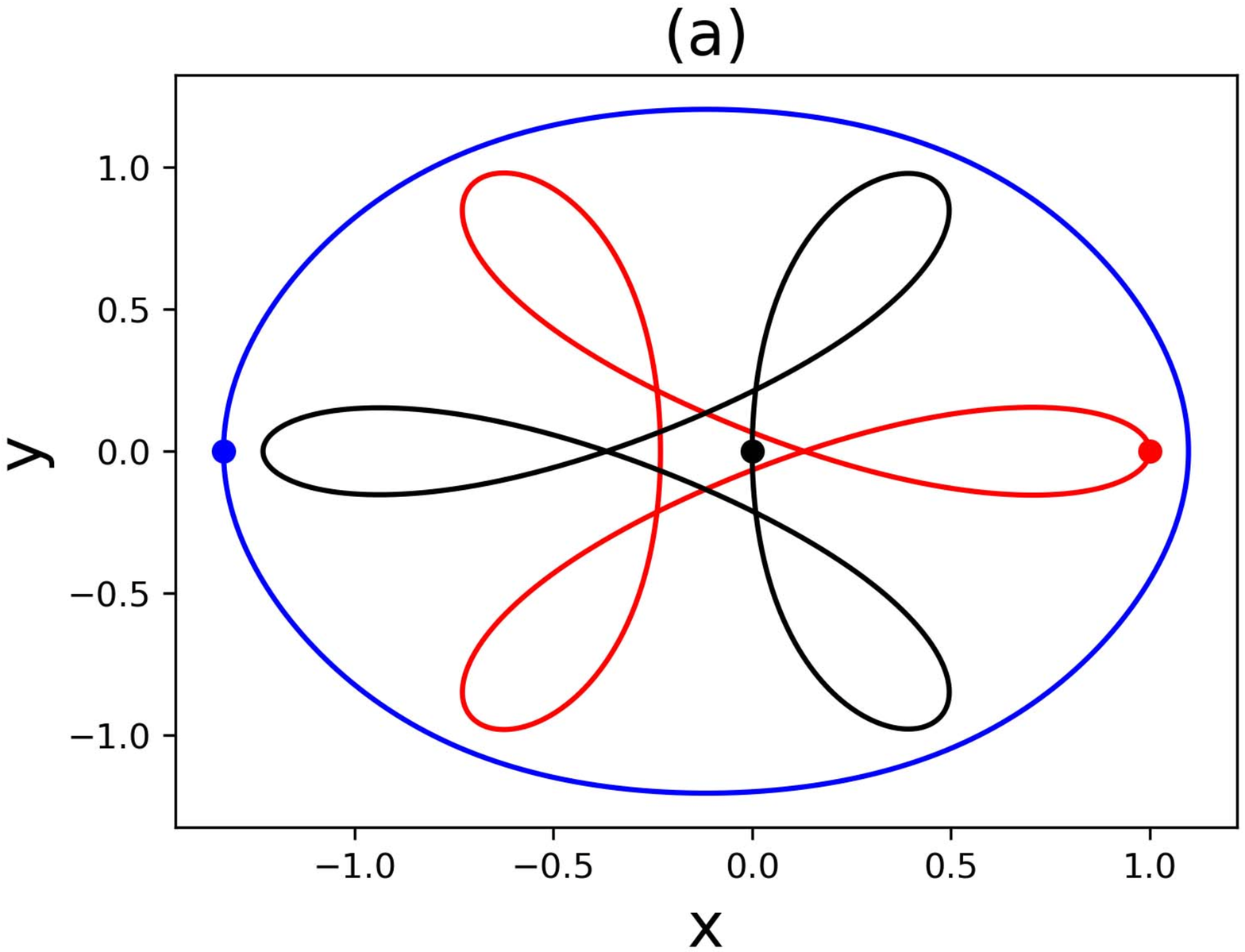}
		\includegraphics[width=4.2cm]{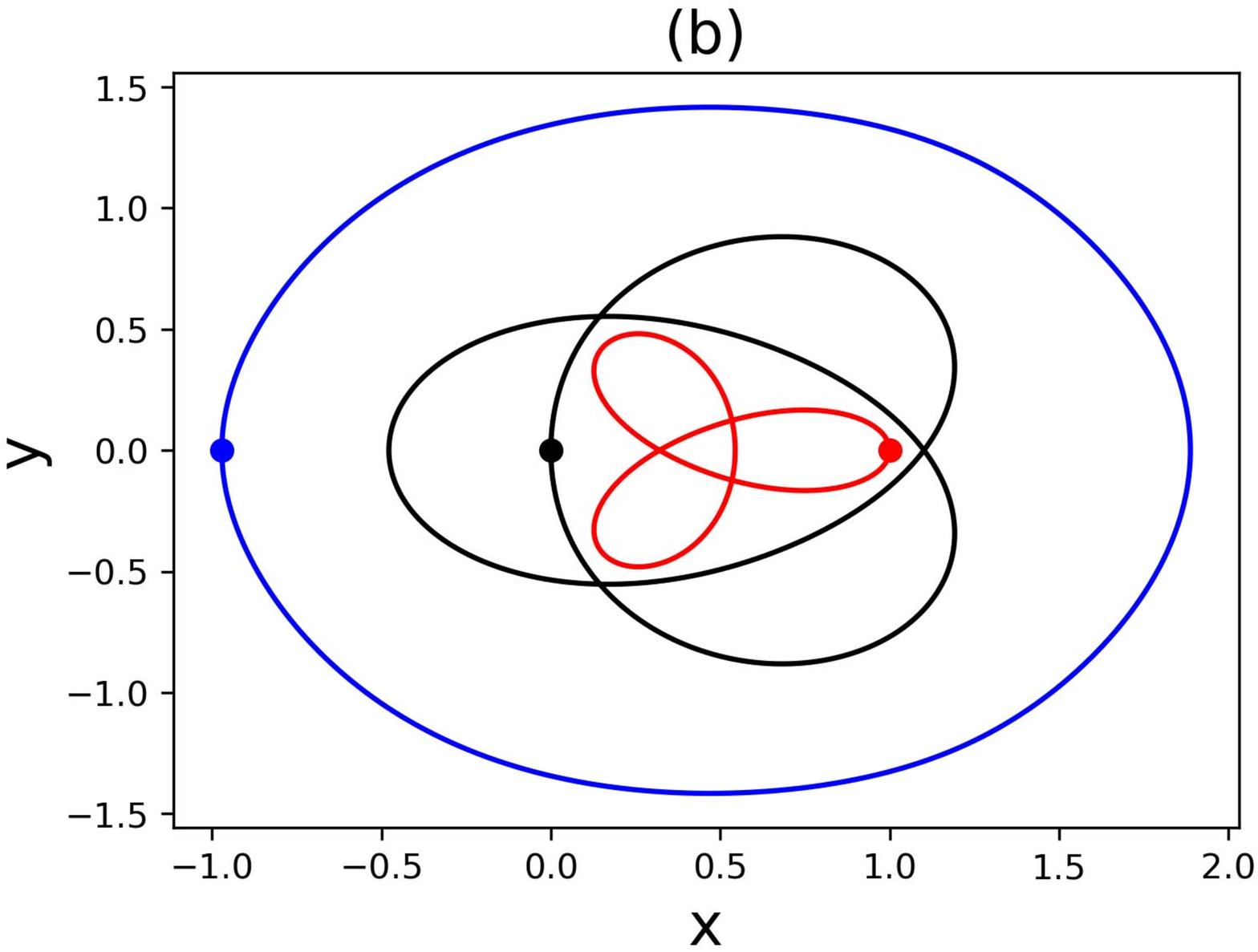}
		\includegraphics[width=4.2cm]{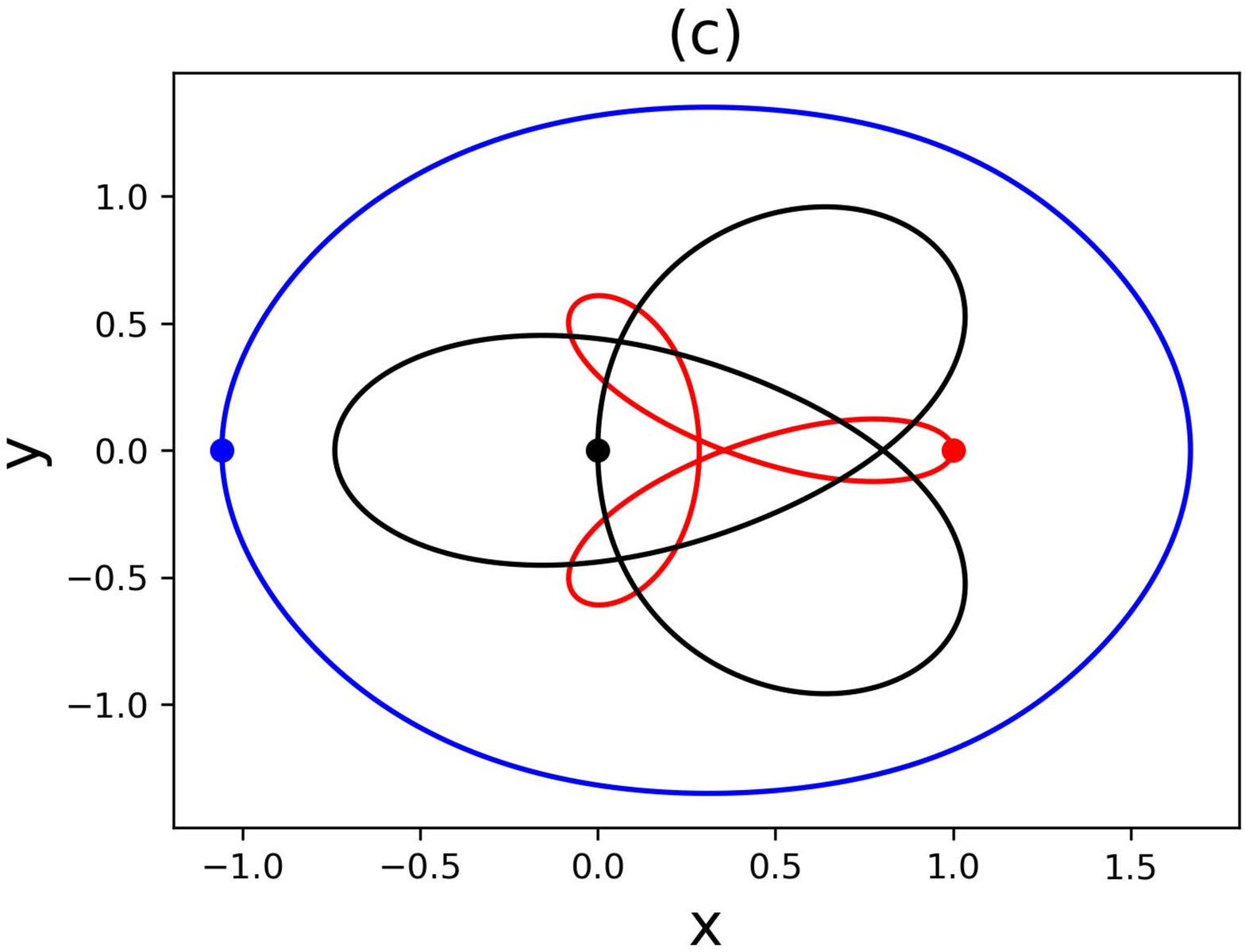}\\
		\includegraphics[width=4.2cm]{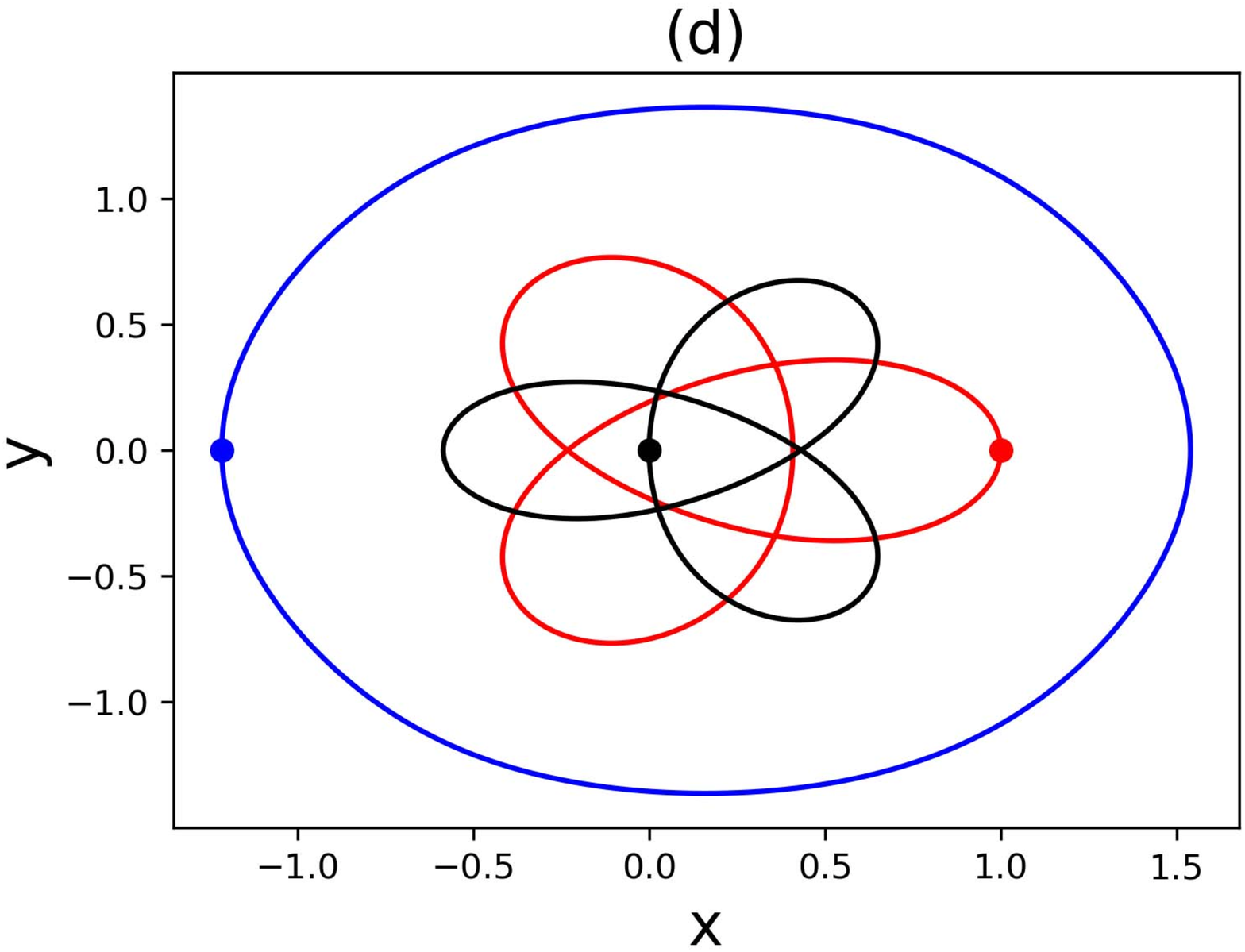}
		\includegraphics[width=4.2cm]{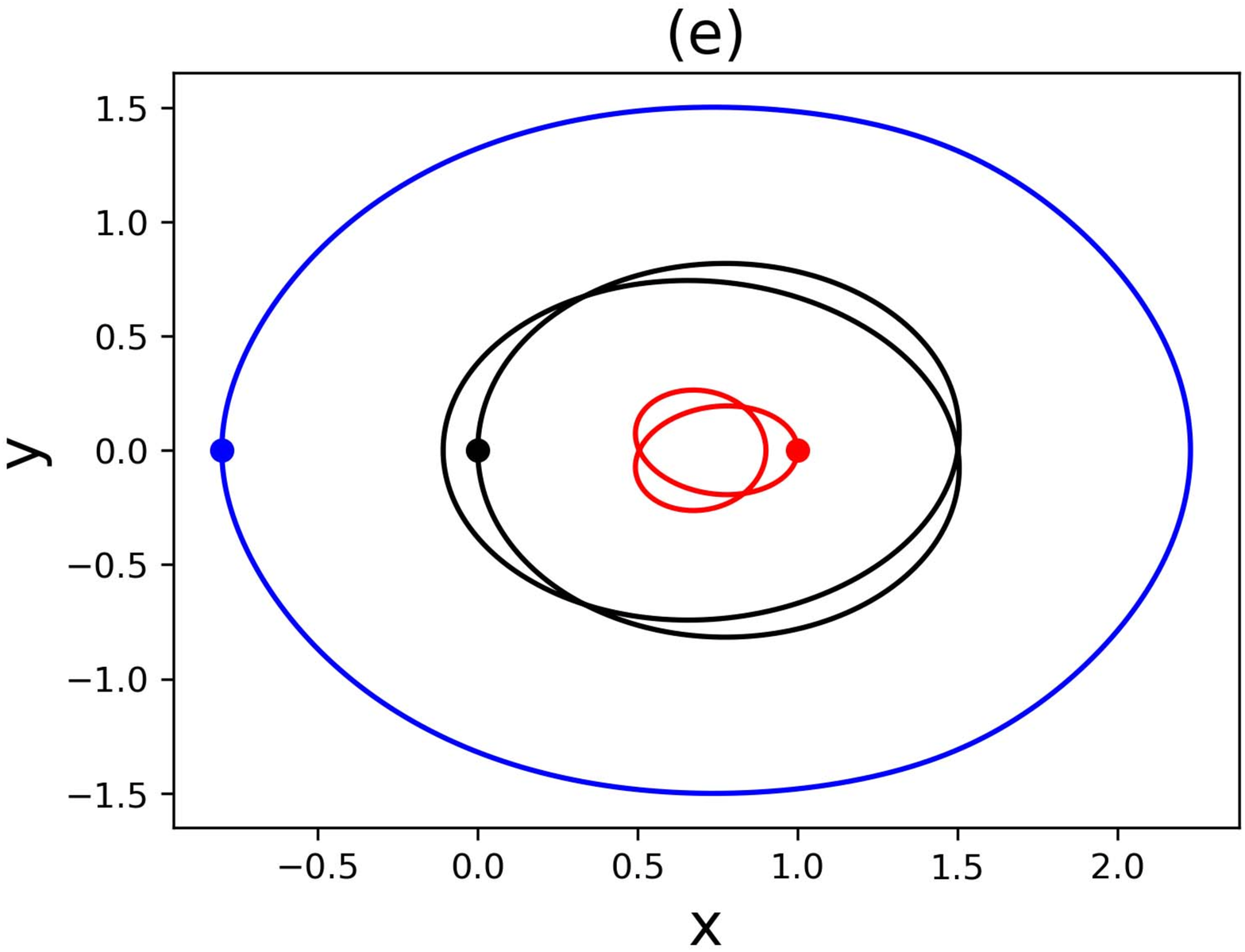}
		\includegraphics[width=4.2cm]{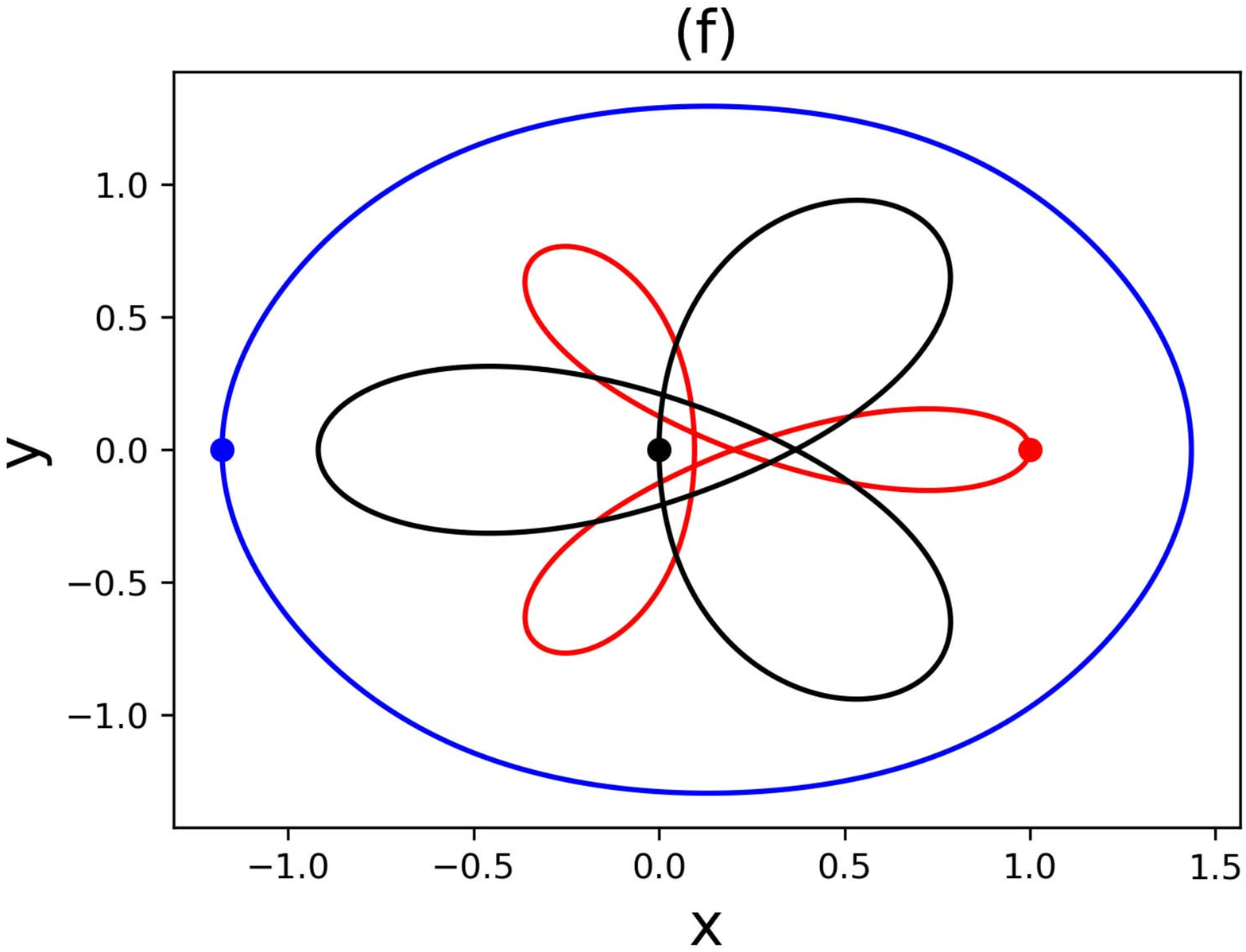}
		\caption{The relatively periodic BHH satellites orbits of the three-body system with various masses $m_1$ and $m_2$ in a rotating frame of reference. The corresponding physical parameters are given by ANN in Table~\ref{predict1}. Blue line: body-1; red line: body-2; black line: body-3.}
		\label{orbit1}
	\end{figure*}
	
			\begin{table*}[t]
		\tabcolsep 0pt \caption{Initial conditions and periods $T$ predicted by our trained ANN model for the six examples of BHH satellites with various masses $m_1$ and $m_2$ in case of  $\bm{r}_1(0)=(x_1,0)$, $\bm{r}_2(0)=(1,0)$, $\bm{r}_3(0)=(0,0)$, $\dot{\bm{r}}_1(0)=(0,v_1)$, $\dot{\bm{r}}_2(0)=(0,v_2)$,  $\dot{\bm{r}}_3(0)=(0,-(m_1v_1+m_2v_2)/m_3)$, where $m_i$, $x_i$ and $v_i$ are the mass, initial position and velocity of the $i$-th body, respectively, with the same rotation angle $\theta = 0.38316088765562$ of the reference frame for relatively periodic orbits.} \small \label{predict1}
		\begin{center}
			\footnotesize
			\def\temptablewidth{1\textwidth}
			{\rule{\temptablewidth}{1pt}}
			\begin{tabular*}{\temptablewidth}{@{\extracolsep{\fill}}cllllll}
				Case & $m_1$ & $m_2$ & $x_1$ & $v_1$  & $v_2$  & $T$\\
				\hline
				(a)	&	1.0124	&	0.9968	& -1.32962	   &	-0.88963           &	-0.28501	         &        9.2111	\\
				(b)	&	0.5312	&	2.2837	& -0.97138            &	-1.37584 	        &	-0.34528            &	  7.0421  	\\
				(c)	&	0.8056	&	2.0394	& -1.05795 	    &	-1.24044	        &	-0.26723            &	  7.4389    \\
				(d)	&	0.3916	&	0.8341	& -1.21503             &	-1.00328	         &	-0.53749            &	  9.2807    \\
				(e)	&	0.1472	&	3.4219	& -0.80027 	     &	-1.74811	         &      -0.42176	         &	  5.8793	\\
				(f)	&	0.8413	&	1.4155	& -1.17777	     &	-1.05903	         &	-0.29934             &	  8.3444  \\
			\end{tabular*}
			{\rule{\temptablewidth}{1pt}}
		\end{center}
	\end{table*}

	The relationship ${\cal F}^*$ between $(m_1, m_2)\in S^*$ and $(x_1, v_1, v_2, T)$ given by the ANN model is fundamentally different from the set of the original 29150 periodic orbits in the following aspects.   
	\begin{enumerate}
		\item[(A)]     For {\em arbitrary} values of $(m_1, m_2)\in S^*$,  ${\cal F}^*$  can always give a good enough prediction of the initial condition $(x_1, v_1, v_2)$ and the period $T$ of the corresponding periodic orbit, which can be used as a starting point to gain a more accurate periodic orbit with a tiny return distance  (deviation) $\delta_T < 10^{-10}$.   		
		\item[(B)]     For each prediction of a periodic orbit given by our ANN model for arbitrary values of $(m_1, m_2)\in S^*$, we can always modify it by means of the  Newton-Raphson method  \cite{Farantos1995, Lara2002,  Abad2011},  until a very accurate periodic orbit with $\delta_T < 10^{-60}$ is gained.   
		For examples,  for the randomly chosen masses $m_1=0.550073$ and $m_2=1.738802$,  the initial condition and period of the periodic orbit predicted by ${\cal F}^*$ is $x_1=-1.0509$, $v_1=-1.2291$,  $v_2=-0.3751$ and $T=7.7189$.   The return distance  $\delta_T$ of this predicted periodic orbit  is  about $6 \times 10^{-4}$, which  can be reduced to $10^{-60}$ by means of the Newton-Raphson method  \cite{Farantos1995, Lara2002,  Abad2011} and the CNS, with the initial condition and period in accuracy of hundred significant digits:  
\begin{eqnarray}
&x_1 = -1.0509175496041811604923698392786632425650672636140394304186656973,  \nonumber \\
&v_1 = -1.2291270518667041014360821708612407200218851479344331903064815068,  \nonumber \\
&v_2 = -0.37510727670051981894849679865106213490791658967297597038682248759, \nonumber \\
&T =  +7.7189475555888051714951891045174205372467192448577195610635970572.
\end{eqnarray}
		Note that the prediction given by our ANN model are in the accuracy of four significant digits.  
		Similarly,  we {\em randomly} check one thousand cases of $(m_1, m_2)\in S^*$,    and found that, in every case,  we indeed can always gain a periodic orbit with a quite tiny return distance  (deviation) $\delta_T < 10^{-60}$.     
		All of these suggest that,  statistically speaking,  for {\em arbitrary} mass values   $(m_1, m_2) \in S^*$,  every prediction given by our ANN model could lead to a periodic orbit that could be in {\em arbitrary} accuracy as long as the performance of computer is good enough.  
		
		\item[(C)] It is found that the mean absolute errors of predictions for randomly selected one thousand cases of  $(m_1, m_2)\in S^*$ given by our trained ANN model ${\cal F}^*$ are in the level of $10^{-5}$ for  the initial conditions $(x_1, v_1, v_2)$ and $10^{-4}$ for the period $T$.   Although we  indeed can further gain a periodic orbit in accuracy of one hundred significant digits by means of the Newton-Raphson method  \cite{Farantos1995, Lara2002,  Abad2011} and the CNS,  the periodic orbits predicted  by our trained ANN model are good enough from practical viewpoint,  since it is {\em unnecessary} to have so accurate trajectory in practice. For example, there also exits error in measurement for the actual astronomical observation.  
		In practice,  it is rather convenient to gain a periodic orbit for {\em arbitrary} masses $(m_1, m_2)\in S^*$ by means of the initial conditions $(x_1, v_1, v_2)$ and period $T$ predicted by the trained ANN model,  for example,  as shown in Figure~\ref{orbit1} for randomly chosen six different masses of $m_1$ and $m_2$.   
		Compared with using the numerical continuation method to find a periodic orbit with arbitrary mass  $(m_1, m_2)\in S^*$, using the neural network can directly provide a good enough prediction of periodic orbit in practice.  While for the numerical continuation method, we need to modify the initial conditions by the Newton-Raphson method to give a good enough solution. Even if we aim to obtain the periodic orbits with high precision, the neural network can give more accurate initial solution for the Newton-Raphson method. Thus, our ANN model  provides  us  great  convenience  in  practice.

		\item[(D)]  The ANN model can ceaselessly learn and thus be further modified when some new periodic orbits are gained, say, the trained ANN model could become wiser and wiser.  
		
	\end{enumerate}

	In summary,  we illustrate that the trained ANN model can provide us  accurate enough periodic orbits for {\em arbitrary} values of $(m_1, m_2)$ in an irregular domain $S^*$.          
	
	How accurate is a periodic orbit with return distance (deviation) $d_{T}<10^{-60}$ ?  Let $\Delta x_{1}$  denote the dimensionless deviation of the initial position.   Even if we use the diameter of universe  $d_{u}$ = 930 light year =  $8.8 \times 10^{18}$ meter as the characteristic length $L$, we have the corresponding inaccuracy of the dimensional initial position $ | \Delta x_{1}| d_{u} < 10^{-41}$ meter, which is  six order of magnitude {\em smaller} than  the Planck length $l_{p}\approx 1.62 \times 10^{-35}$ meter.   Note that Planck length is a lower bound to physical proper length in any space-time:  it is {\em impossible} to measure length scales smaller than the Planck length, according to Padmanabhan  \cite{Padmanabhan1985}.    Besides,  it should be emphasized that all of these periodic orbits are {\em stable}, say, a tiny disturbance does not increase exponentially.    So, from physical view-point,  a stable  periodic  orbit  with return distance (deviation)  $d_{T}<10^{-60}$ gained by numerical method is physically  {\em equivalent} to  $d_{T}= 0$ that corresponds to an {\em exact} solution of periodic orbit in a closed-form.  
	
	Therefore, the high-performance computer and the machine learning play very important role in finding periodic orbits of three-body systems with arbitrary masses.   It should be emphasized here that it is Turing  \cite{Turing1936, Turing1950} who laid the foundations of  modern  computer and artificial intelligence (including machine learning).

	\begin{figure}[t]
		\centering
		\includegraphics[width=8.5cm]{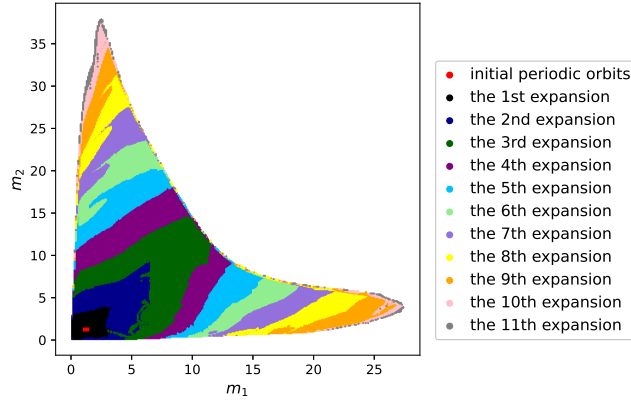}
		\caption{The relatively periodic orbits with the same rotation angle $\theta = 0.0105056462558377$ of reference frame, found in each extrapolation/expansion on the various mass regions. Red dot: initial periodic orbits; black dot: 1st expansion; dark blue dot: 2nd expansion; dark green dot: the 3rd expansion. dark purple dot: the 4th expansion; light blue dot: the 5th expansion; light green dot: the 6th expansion; light purple dot: the 7th expansion; yellow dot: the 8th expansion; orange dot: the 9th expansion; pink dot: the 10th expansion; grey dot: the 11th expansion.}
		\label{expand-arc2}
	\end{figure}

	\begin{figure}[t]
		\centering
		\includegraphics[width=4.2cm]{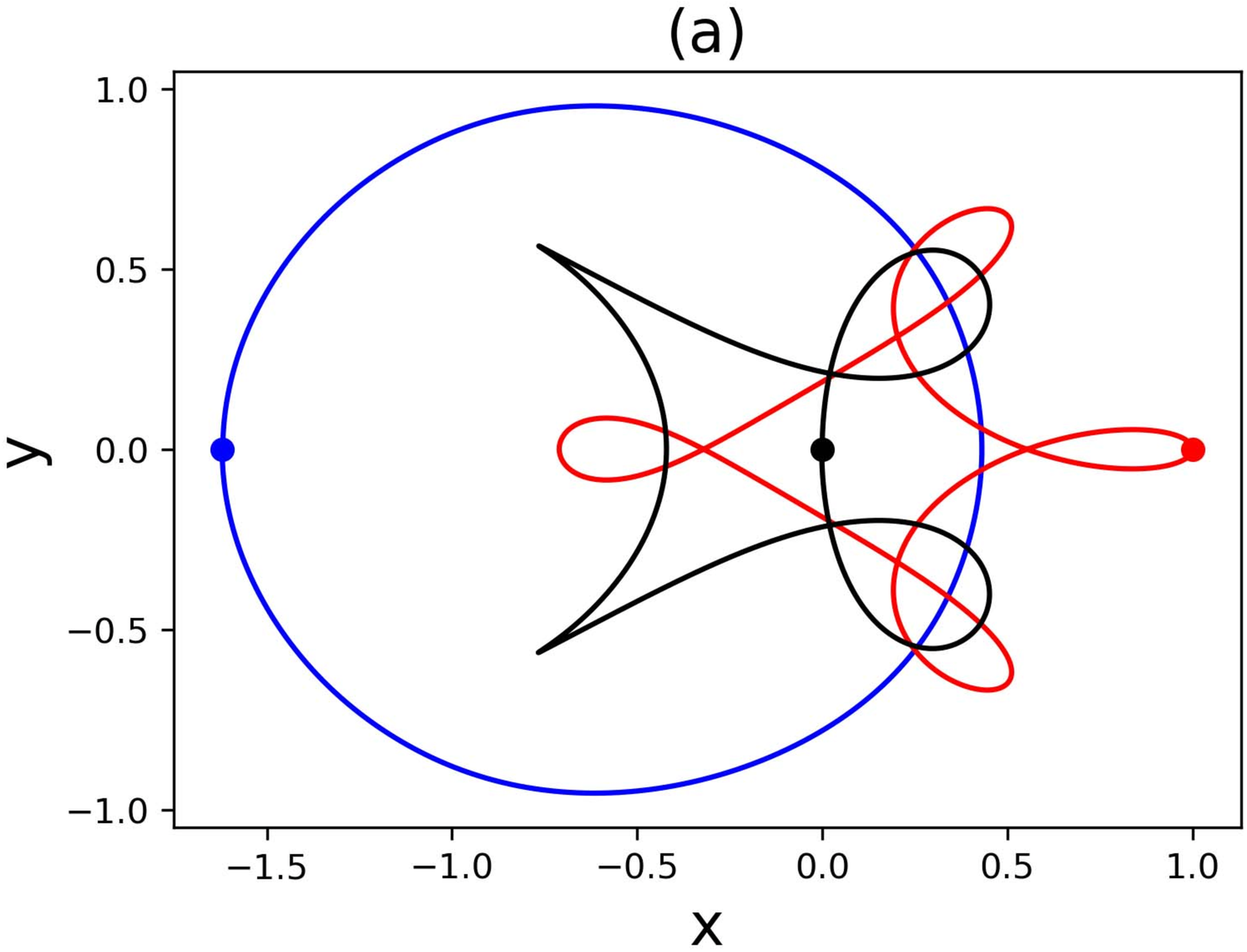}
		\includegraphics[width=4.2cm]{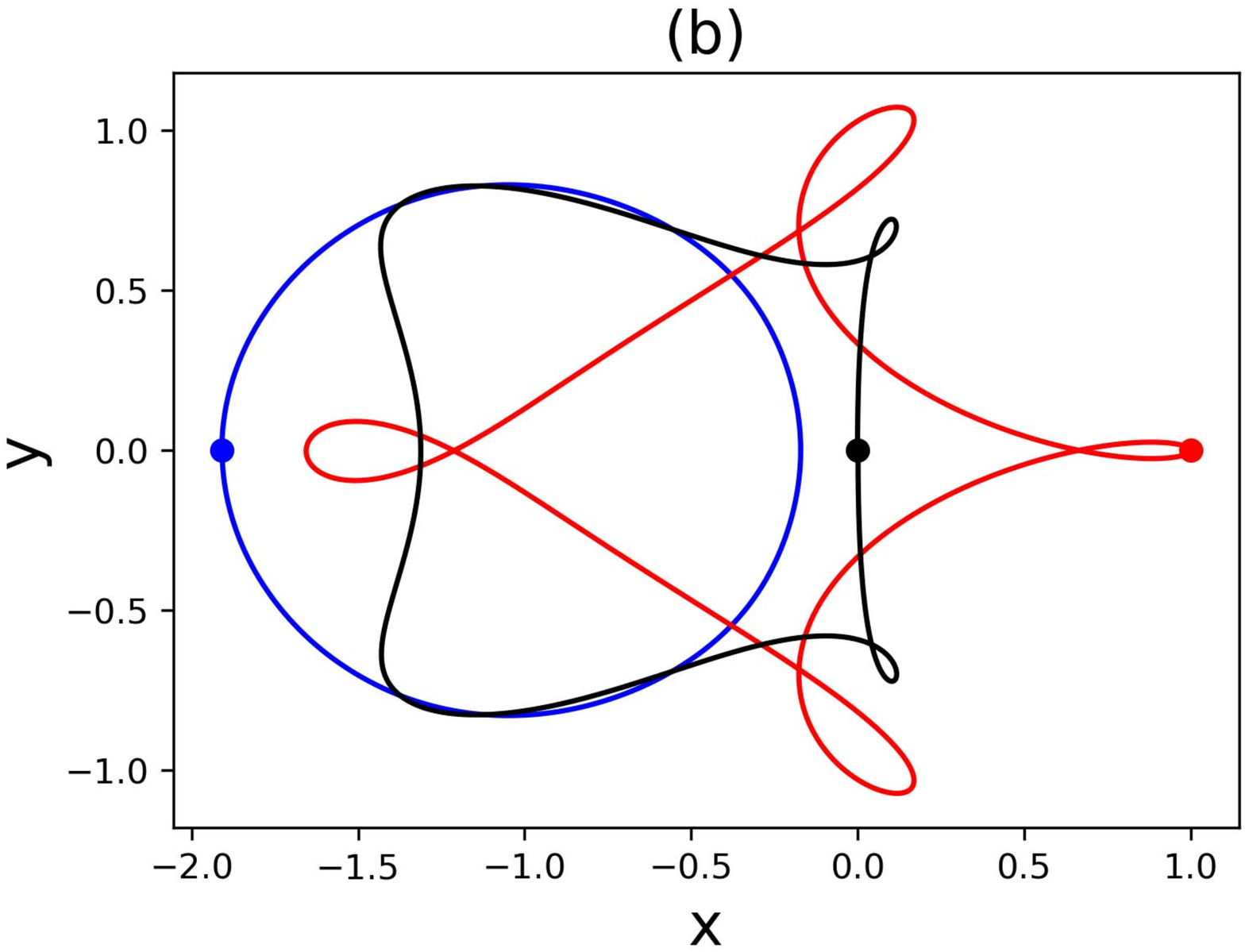}
		\includegraphics[width=4.2cm]{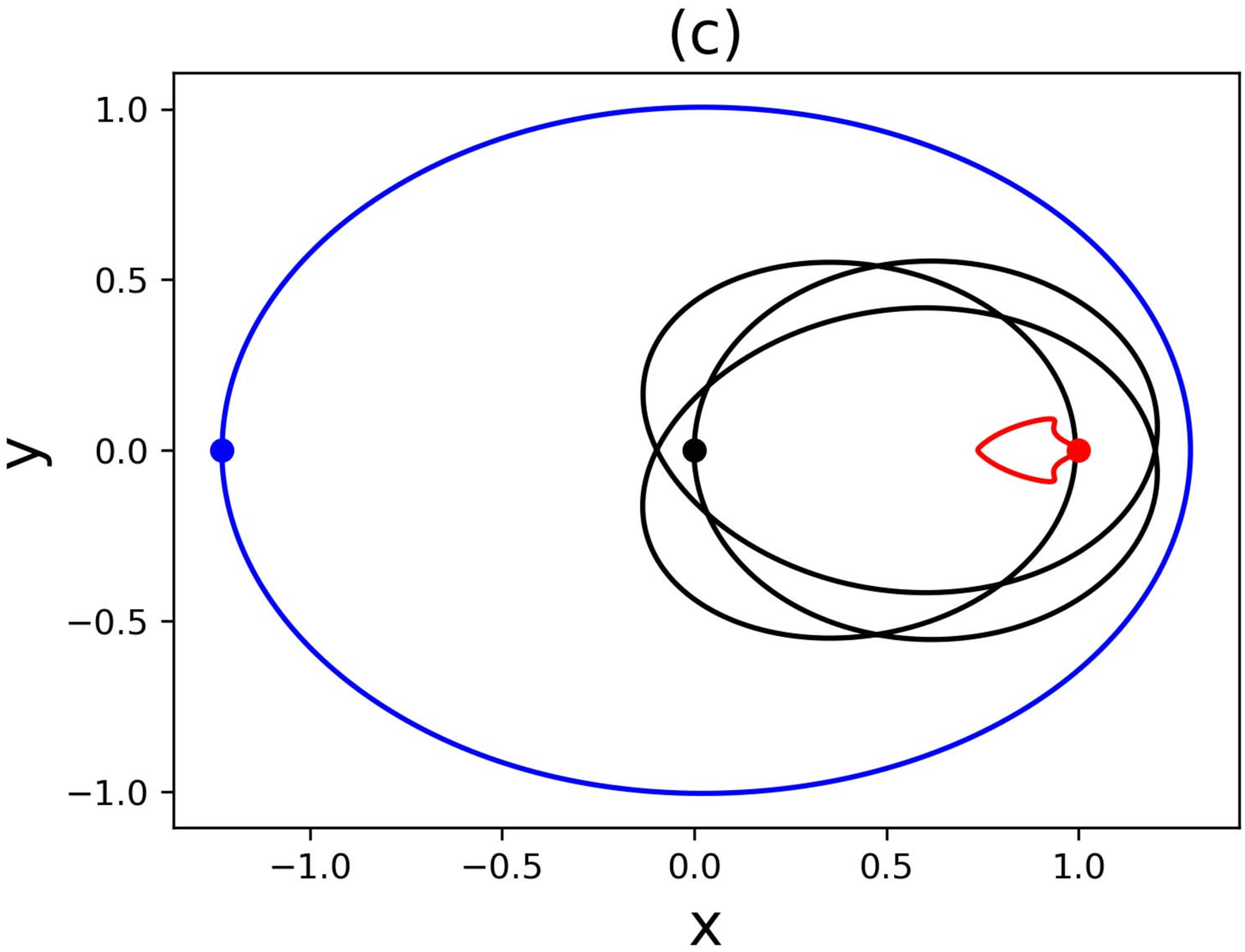}\\
		\includegraphics[width=4.2cm]{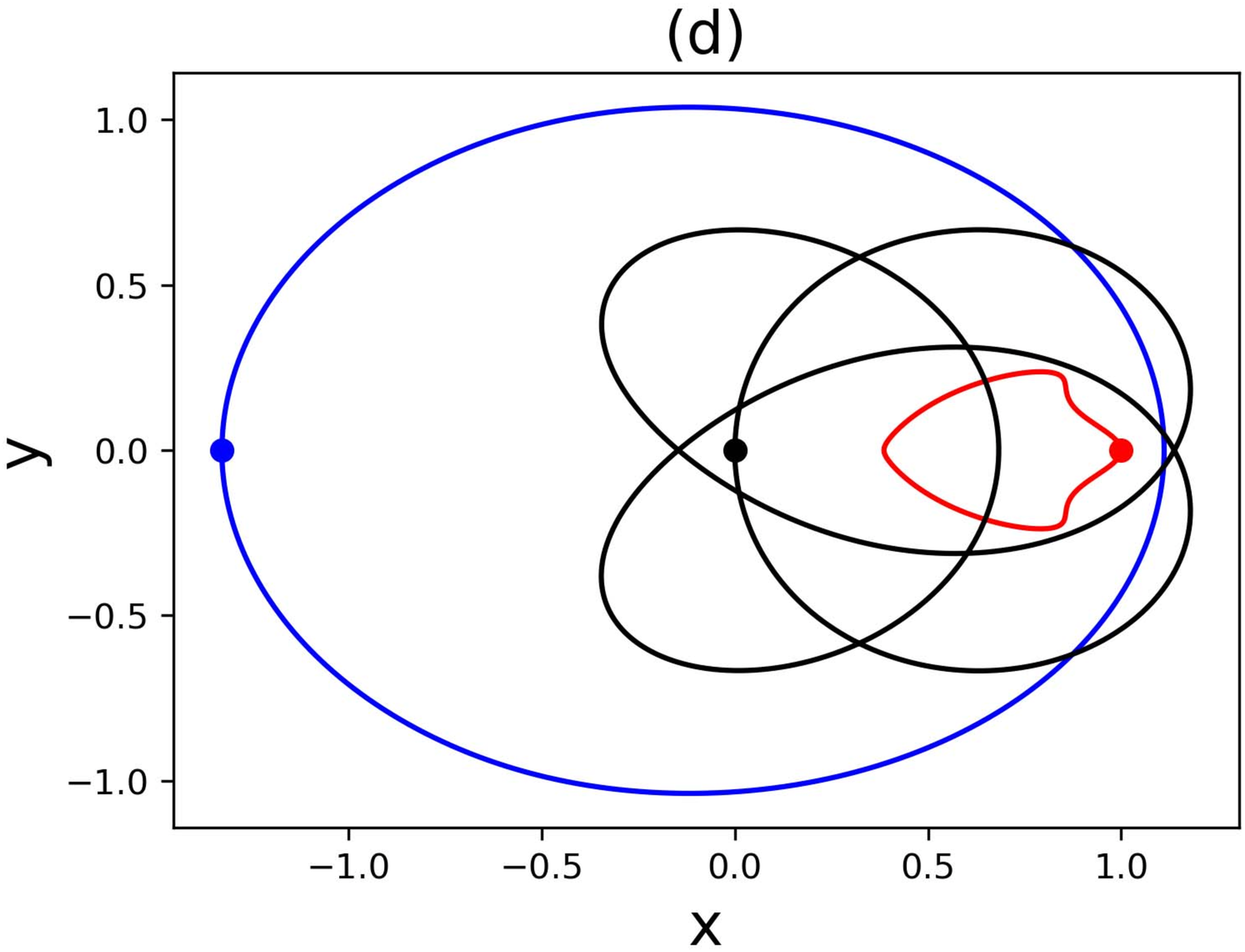}
		\includegraphics[width=4.2cm]{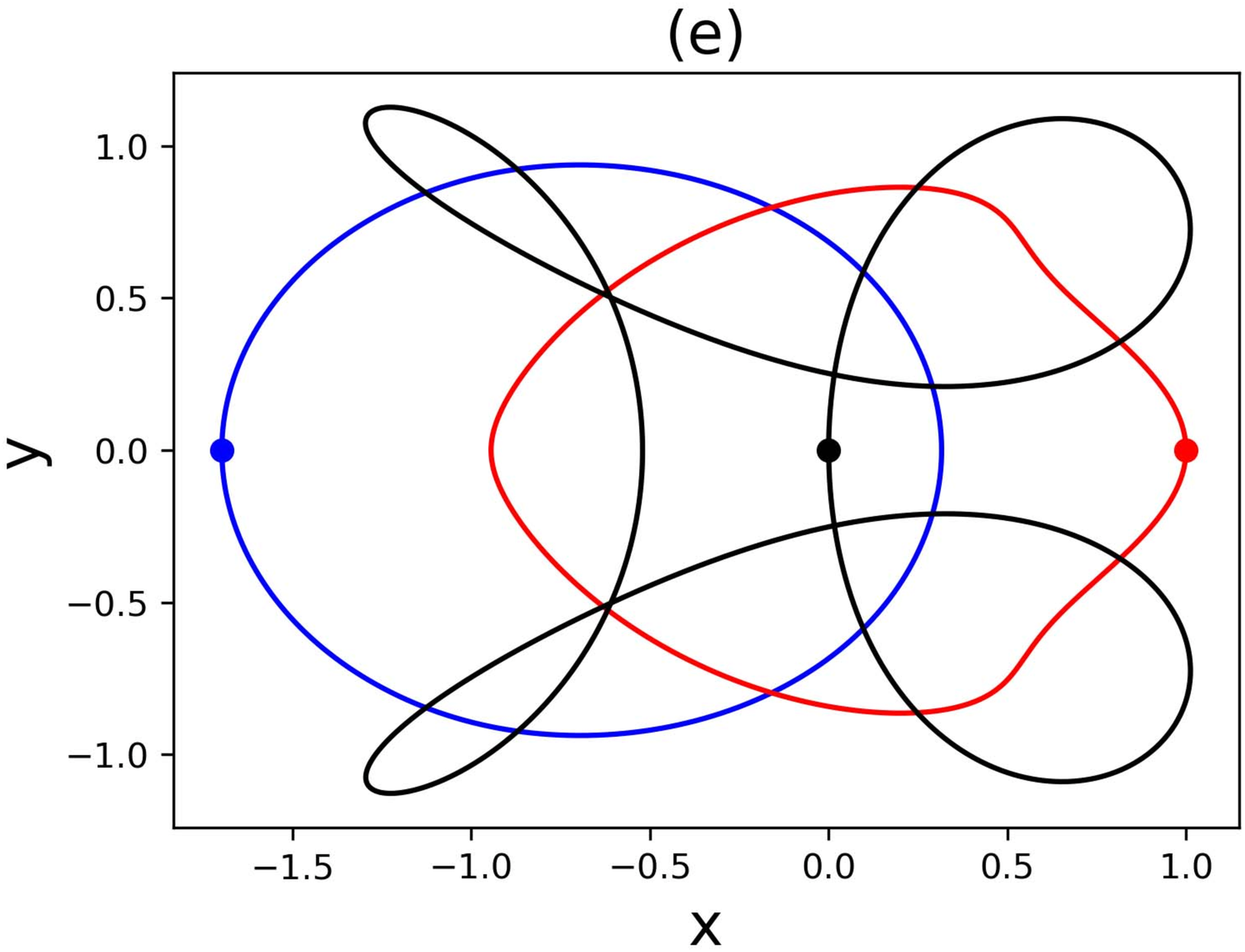}
		\includegraphics[width=4.2cm]{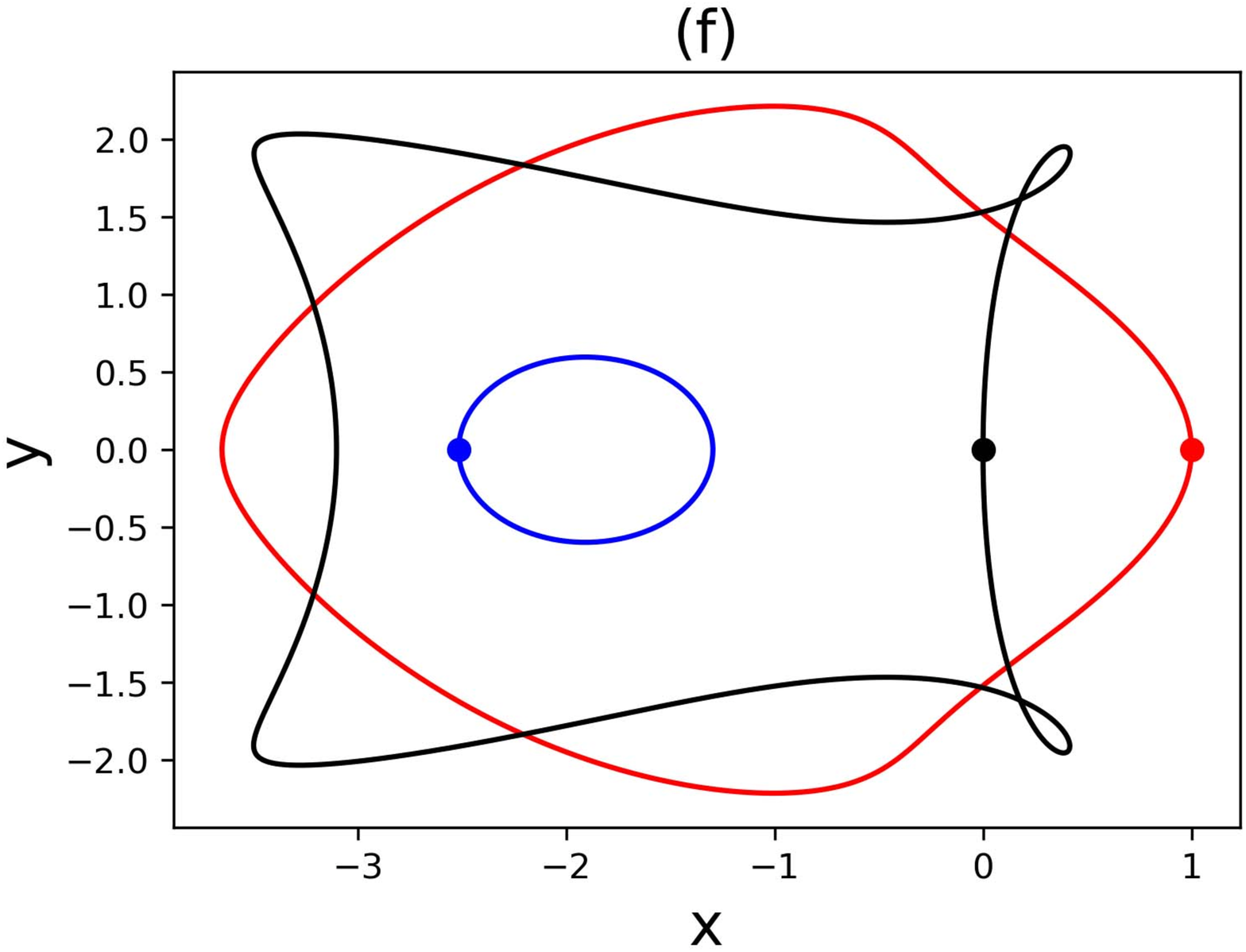}
		\caption{The relatively periodic BHH satellites orbits of the three-body system with various masses $m_1$ and $m_2$ in a rotating frame of reference. The corresponding physical parameters are given by ANN in Table~\ref{predict2}. Blue line: body-1; red line: body-2; black line: body-3.}
		\label{orbit2}
	\end{figure}
	
	The above approach has general meaning. To show this point, let us further consider another BHH satellite periodic orbit with three equal masses $m_1 = m_2=m_3=1$, $x_1$ = $-1.609965115714630$, $v_1=-0.6656909425824538$, $v_2= -0.1529561125709906$, the period $T=6.879203007710456$ and the rotation angel $\theta = 0.0105056462558377$ of the reference frame (for relatively periodic orbits).     
	At first, using this known periodic orbit as a starting point, we can obtain 36 periodic orbits with various masses in a small mass domain  $m_1\in [1.0, 1.5]$ and $m_2 \in [1.0, 1.5]$ (with mass increment $\Delta m_1 = \Delta m_2 = 0.1$) by combining  the numerical continuation method  \cite{Allgower2003} and the Newton-Raphson method  \cite{Farantos1995, Lara2002,  Abad2011}.  
	
	Similarly, we train a ANN model (with the same structure as mentioned above) by means of the initial conditions and periods of these 36 known periodic orbits and then use the trained ANN model to predict the initial conditions and periods of some  candidates of possible periodic orbits {\em outside} the previous mass domain, while the initial conditions and periods of each candidate are modified  by the Newton-Raphson method  \cite{Farantos1995, Lara2002,  Abad2011} so as to confirm whether or not the candidate is a periodic one, say, its return distance can be reduced to the tiny level of $10^{-10}$.  The same process repeats 11 times,  while the number of the known relatively periodic orbits becomes more and more, the mass domain becomes larger and larger,  and the ANN model becomes wiser and wiser!   Finally,  we totally obtain 35895 relatively periodic orbits in a mass domain $\bar{S}^*$ whose area is about 1000 times larger than the initial one $m_1\in [1.0, 1.5]$ and $m_2 \in [1.0, 1.5]$, as shown in Fig.~\ref{expand-arc2}.

	\begin{table*}[t]
		\tabcolsep 0pt \caption{Initial conditions and periods $T$ of six examples of BHH satellites with various masses $m_1$ and $m_2$, predicted by our trained ANN model in case of  $\bm{r}_1(0)=(x_1,0)$, $\bm{r}_2(0)=(1,0)$, $\bm{r}_3(0)=(0,0)$, $\dot{\bm{r}}_1(0)=(0,v_1)$, $\dot{\bm{r}}_2(0)=(0,v_2)$,  $\dot{\bm{r}}_3(0)=(0,-(m_1v_1+m_2v_2)/m_3)$, where $m_i$, $x_i$ and $v_i$ are the mass, initial position and velocity of the $i$-th body, respectively, with the same rotation angle $\theta = 0.0105056462558377$ of the reference frame for relatively periodic orbits.  } \small \label{predict2}
		\begin{center}
			\footnotesize
			\def\temptablewidth{1\textwidth}
			{\rule{\temptablewidth}{1pt}}
			\begin{tabular*}{\temptablewidth}{@{\extracolsep{\fill}}cllllll}
				No. & $m_1$ & $m_2$ & $x_1$ & $v_1$  & $v_2$  & $T$\\
				\hline
				(a)	&	$1.0283$	&	$0.9879$	& $-1.62064$	       &	$-0.65955$           &	$-0.14784$	           &   $6.9193$	\\
				(b)	&	$1.5142$	&	$0.4968$	& $-1.90809$            &	$-0.50283$ 	            &	$-0.08614$           &	  $8.2924$ 	\\
				(c)	&	$2.9216$	&$31.9067$	& $-1.22892$	       &	$-2.37239$	            &	$0.09481$           &	 $1.6822$\\
				(d)	&	$4.4143$	&	$18.6575$	& $-1.32880$ 	       &	$-1.86103$	             &$0.25507$	           &	 $2.3306$	\\
				(e)	&	$10.3501$	&	$10.4522$	& $-1.69797$             &	$-1.22777$	            &	$0.86183$           &	$3.5607$ \\
				(f)	&	$18.7011$	&	$4.2388$	& $-2.51585$	       &	$-0.53262$	            &	$1.49754$          &	 $5.9612$ \\
				
			\end{tabular*}
			{\rule{\temptablewidth}{1pt}}
		\end{center}
	\end{table*}

	Similarly, we further train our ANN model by using the known 35895 relatively periodic orbits.  To show the validation and accuracy of this ANN's model, we  randomly chose 1000 cases of {\em arbitrary} mass $(m_1, m_2)\in \bar{S}^*$, and it is found that the initial conditions and periods predicted by our trained ANN model  can  provide us good enough periodic orbits with a return distance $d_{T}$ in a tiny level of $10^{-3}$ (about 71$\%$), $10^{-4}$ (about  2$\%$), and $10^{-2}$ (about 27$\%$), which can be further decreased  by means of the clean numerical simulation (CNS) and the Newton-Raphson method with multiple-precision (MP) until a very small return distance (such as $d_{T} < 10^{-60}$) is satisfied.  Thus, generally speaking,  our trained ANN model can always give periodic orbits for an arbitrary mass $(m_1, m_2)\in S_2^*$, which  is accurate enough from practical viewpoint.  For example, randomly choosing six cases with various masses $m_1$ and $m_2$, where $(m_1, m_2)\in \bar{S}^*$,  our ANN model can quickly predict  their  initial conditions and periods,  as listed in Table~\ref{predict2}, and give the corresponding trajectories in a satisfied level of accuracy, as shown in Fig~\ref{orbit2}.  It is found that the differences between the periodic orbits predicted by the ANN's model with $10^{-4} < d_{T}<10^{-2}$ and the ``exact'' ones  further modified by the Newton-Raphson method and the multiple-precision (MP) with $d_{T} < 10^{-60}$ are negligible for practical uses.   Thus, our approach based on the ANN model has indeed general meanings.

	\section{Classification for orbits based on machine learning}

\begin{figure}[t]
		\centering
		\includegraphics[width=7cm]{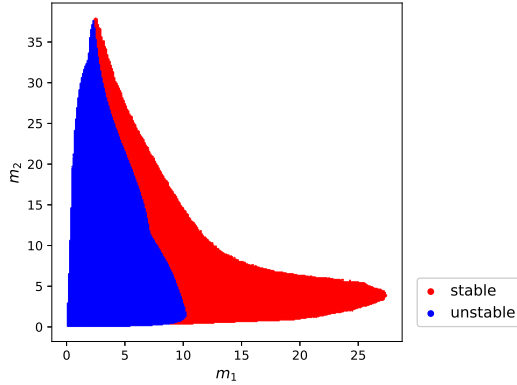}
		\caption{The linear stability of the relatively periodic orbits in the second case with the rotation angle $\theta = 0.0105056462558377$. Red domain: stable; blue domain: unstable.}
		\label{stable}
	\end{figure}

\begin{figure}[t]
		\centering
	    \subfigure[]{\includegraphics[width=6.5cm]{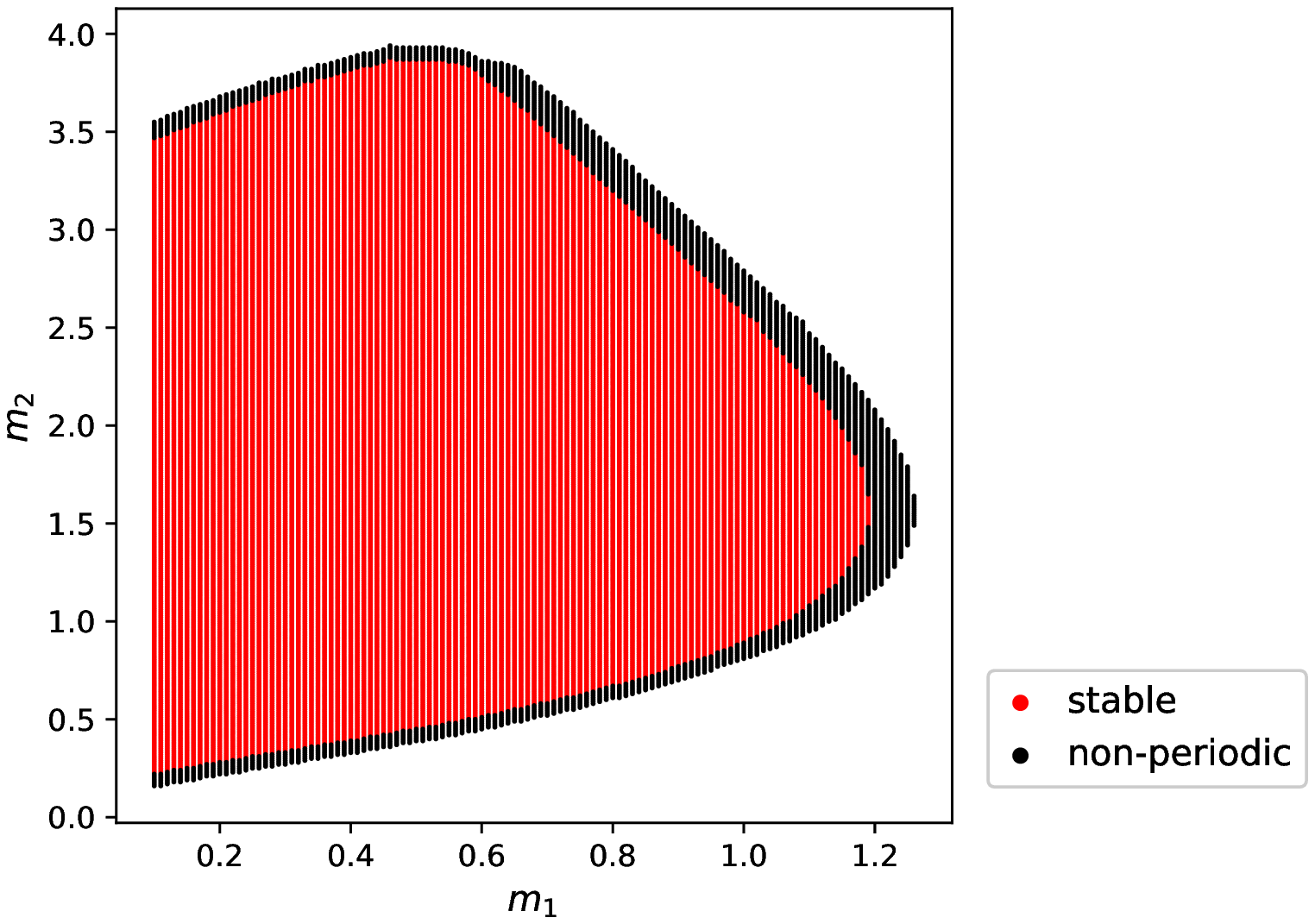}}
		\subfigure[]{\includegraphics[width=6.5cm]{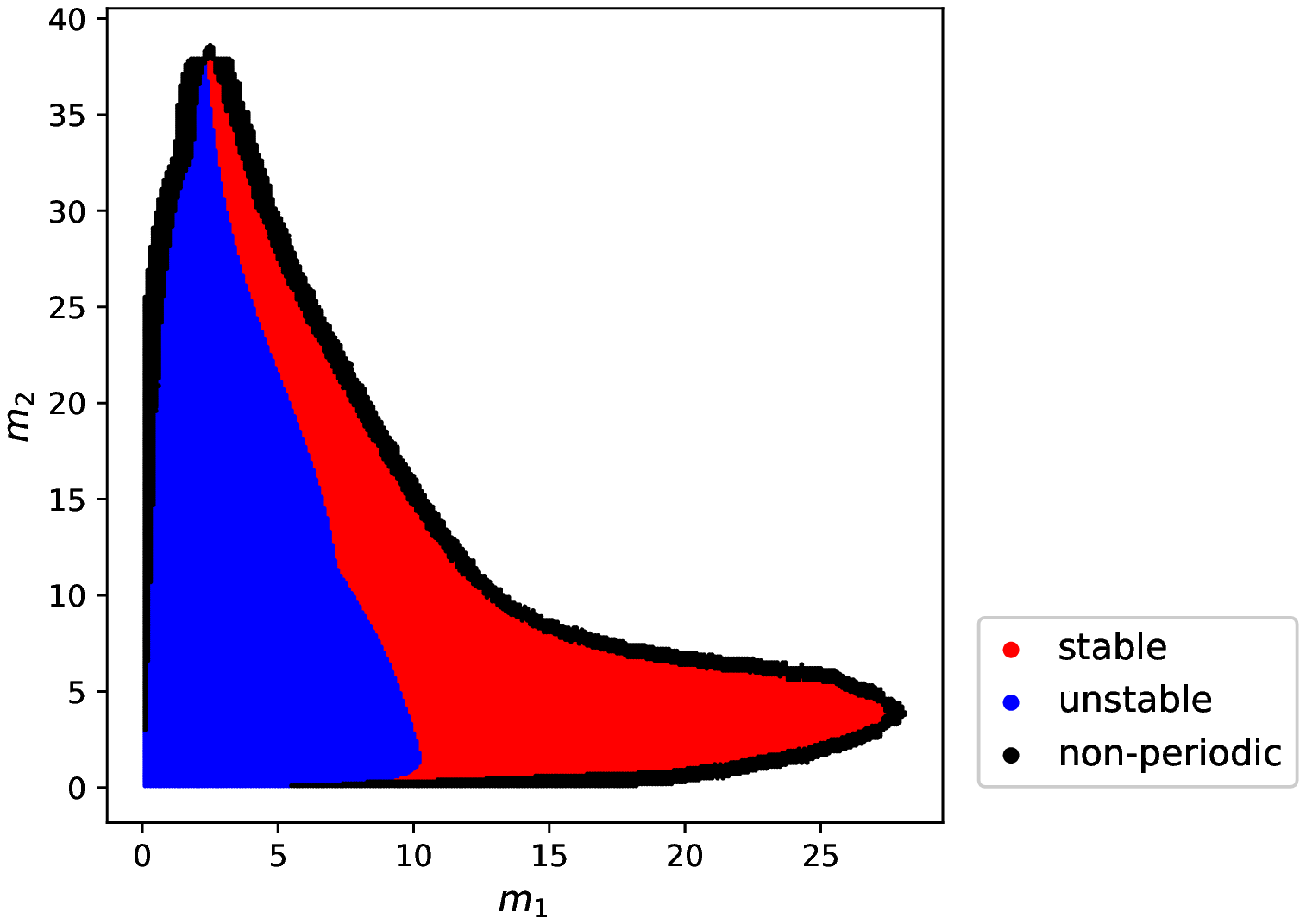}}
		\vspace{0.1cm}
		\caption{The linear stability and periodicity of the orbits for the first case (a) and the second case (b). Red dot: linear stable; blue dot: unstable; black dot: non-periodic.}
		\label{stable2}
	\end{figure}

		Stability is an important property of periodic orbits, because only stable triple systems can be observed.  We employ a theorem given by Kepela and Sim\'o  \cite{kapela2007} to determine the linear stability of periodic orbits of three-body problem through the monondromy matrix   \cite{Simo2002}.  It is found that all relatively periodic orbits for the first case with the rotation angle $\theta = 0.3831608876556280$ are linearly stable.  For the second case with the rotation angle $\theta = 0.0105056462558377$,   there are 16739 linearly stable periodic orbits among the 35895 computer-generated  periodic orbits,  as shown in Fig~\ref{stable}.  Note that,  the ANN can be applicable to deal with complicated classification problems   \cite{bala2017classification}. We use the ANN to classify different types of orbits for different masses which have complex shapes of the boundary.  We can give an ANN classifier model to classify the periodicity and stability of the orbits for {\em arbitrary}  masses $m_{1}$ and $m_{2}$, especially for the masses nearly the boundary of each type.    The orbits are classified into three categories: stable periodic orbits, unstable periodic orbits and non-periodic orbits,  expressed by  $[1, 0, 0]$, $[0, 1, 0]$ and $[0, 0, 1]$, respectively.  The ANN classifier model consists of eight fully connected layers with one input layer, six hidden layers and one output layer. The numbers of neurons for input layer, hidden layers and output layer are 2, 256 and 3, respectively.  Different from the above-mentioned ANN regression model, the activation function of the ANN classifier model is softmax function.  The loss function is cross entropy. As shown in Fig~\ref{stable2}, for the first case, the mass domain of 3128 non-periodic orbits are outside the boundary of periodic orbits. The numbers of stable periodic orbits, unstable periodic orbits and non-periodic orbits are  29150,  0 and 3128, respectively. For the second case, the mass domain of 5900 non-periodic orbits are outside the boundary of periodic orbits. The numbers of stable periodic orbits, unstable periodic orbits and non-periodic orbits are 16739,  19156 and 5900, respectively. For each case, the whole dataset is randomly divided into three sets, the training set (90\%), validation set (5\%) and test set (5\%).
		
	\begin{table*}[t]
		\tabcolsep 0pt \caption{Confusion matrix for the ANN predictions in test set of the first case.} \small \label{matrix1} 
		\begin{center}
			\footnotesize
			\def\temptablewidth{1\textwidth}
			{\rule{\temptablewidth}{1pt}}
			\begin{tabular*}{\temptablewidth}{@{\extracolsep{\fill}}lccc}
				Type & Stable periodic orbits  & Non-periodic orbits  \\
				\hline
				Stable periodic orbits	&	1475	&	2		        	\\
				Non-periodic orbits	&	0	&	137	       	  	\\
	
			\end{tabular*}
			{\rule{\temptablewidth}{1pt}}
		\end{center}
	\end{table*}
	
	\begin{table*}[th]
		\tabcolsep 0pt \caption{ Confusion matrix for the ANN predictions in test set of the second case.} \small \label{matrix2}
		\begin{center}
			\footnotesize
			\def\temptablewidth{1\textwidth}
			{\rule{\temptablewidth}{1pt}}
			\begin{tabular*}{\temptablewidth}{@{\extracolsep{\fill}}lccc}
				Type & Stable periodic orbits  & Unstable periodic orbits & Non-periodic orbits \\
				\hline
				Stable periodic orbits	&	846	&	1	& 5        	\\
				Unstable periodic orbits	&	1	&	943	& 2           	  	\\
				Non-periodic orbits	&	1	&	0	& 291           	  	\\
	
			\end{tabular*}
			{\rule{\temptablewidth}{1pt}}
		\end{center}
	\end{table*}

	The early stopping   \cite{prechelt1998early} is used when training, and we save the ANN model with the maximum accuracy in the validation set. The test set is used to verify the generalization capability of the ANN model.  For the first case, the accuracies of the training set, validation set and test set are about $99.98\%$, $99.94\%$ and $99.94\%$ , respectively. And the macro F1-scores   \cite{Dimitrov2017} of the training set, validation set and test set are about 0.9995, 0.9983 and 0.9980, respectively.  For the second case, the accuracies of the training set, validation set and test set are about $99.62\%$,  $99.57\%$ and $99.52\%$, respectively.  And the macro F1-scores of the training set, validation set and test set are about 0.9941, 0.9935 and 0.9932, respectively.  As the accuracies and F1-scores of the training set and test set are close, overfitting doesn't exist in these two cases. The confusion matrixes for the ANN predictions in the test sets of the first and second case are as shown in Table~\ref{matrix1} and \ref{matrix2}, respectively, which illustrate the good performances of the ANN. 
 Therefore, we can use the ANN classifier to predict the periodicity and stability of orbits for any given masses in each case. 
 
	\section{A roadmap of searching for periodic orbits of three-body problem}
	
	The successful examples mentioned above  suggest  us  a general  road map  for  finding new periodic orbits of three-body system (in case of $m_3=1$) with the same ``free group element'' (word) given by Montgomery's  topological method  \cite{Montgomery1998}:
	\begin{enumerate}
		\item[(1)]  For a three-body system with three or two equal masses,  first find candidates of the initial conditions for possible periodic orbits by means  of  the  grid  search  method, and then modify these candidates  by means of the Newton-Raphson method  \cite{Farantos1995, Lara2002,  Abad2011}, until a satisfied periodic orbit with  a  tiny  enough  return  distance (deviation)  is obtained;   
		
		\item[(2)]  Given a known periodic orbit,  use it as a starting point to gain a few of new periodic orbits with various masses in a small  domain of mass $(m_1,m_2)$ by combining the numerical continuation method  \cite{Allgower2003} and the Newton-Raphson method  \cite{Farantos1995, Lara2002,  Abad2011}.  The initial conditions and periods of all these known periodic orbits form a training set for a ANN model.    
		
		\item[(3)] For a given training set in a mass domain of $(m_1, m_2)$,  train the ANN model to predict initial conditions and periods so that some new periodic orbits {\em outside} of the previous mass domain of $(m_1, m_2)$ could be found by modifying these predictions via the Newton-Raphson method  \cite{Farantos1995, Lara2002,  Abad2011}.   Then, combining the results of these new periodic orbits  with the previous training set, we further have a new training set with more elements, which could further provide us some new periodic orbits in a even larger mass domain of $(m_1, m_2)$ in a similar way, {\em outside} of the previous one.   The same process can repeat again and again, so that more and more periodic orbits are found in a larger and larger mass domain of $(m_1, m_2)$, and the ANN model becomes wiser and wiser, until quit few or no new periodic orbits can be found in a larger domain of $(m_1, m_2)$.   Finally, we have a trained ANN model ${\cal F}^*$ of all periodic orbits in the final mass domain $(m_1, m_2)\in S^*$.   
		
		\item[(4)]  Randomly choose hundreds or thousands of {\em arbitrary} masses  $(m_1, m_2) \in S^*$.  For each case, check whether or not  the trained ANN model ${\cal F}^*$ could give an accurate enough prediction, and in addition whether or not the corresponding  return distance (deviation) could be indeed reduced to $10^{-60}$.  If yes, then the ANN model ${\cal F}^*$ can provide a good enough prediction of periodic orbits in the domain $(m_1, m_2) \in S^*$, from practical viewpoint.        
	\end{enumerate}

	\begin{figure}[t]
	\centering
	\includegraphics[width=8cm]{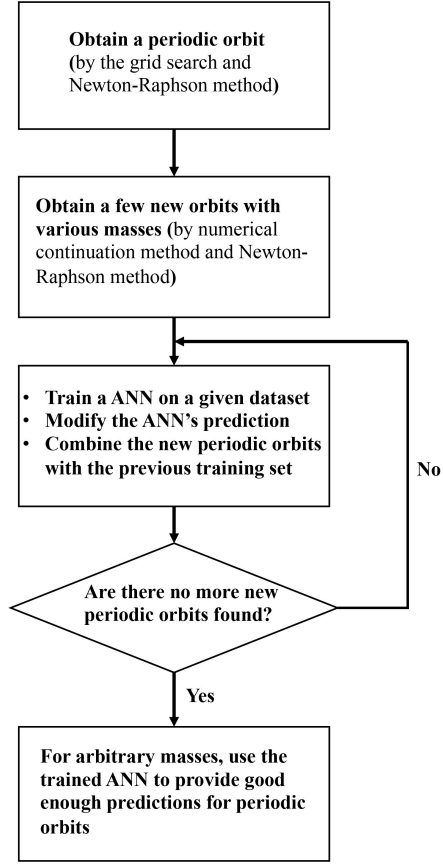}
	\caption{A roadmap for searching the periodic orbits of three-body problem}
	\label{roadmap}
\end{figure}	
	
	The pipeline for searching the periodic orbits of three-body problem is shown in Fig.~\ref{roadmap}. Note that it is better to use the CNS as integrator so as to guarantee the reliability and convergence of computer-generated  trajectories of chaotic three-body system under arbitrary initial conditions in a required long interval of time.  Besides,   the periodicity and stability of orbits for {\em arbitrary} masses $m_{1}$ and $m_{2}$ can be well predicted by an ANN classifier model.

	\section{Concluding remarks and discussions}
	
	The famous three-body problem can be traced back to Newton in 1687, but quite few families of periodic orbits were found in 300 years thereafter.   As proved by Poincar\`{e},  the first integral does not exist for three-body systems, which implies that numerical approach had to be used in general. 
	
	Artificial neural network   \cite{hassoun1995fundamentals, gevrey2003review, andrews1995survey, abiodun2018state} is a machine learning technique evolved from simulating the human brain, which has been widely proved to be rather powerful. Compared with the traditional regression approaches, the ANN has many advantages.  First, the ANN does not require information about the complex nature of the underlying process to be explicitly expressed in mathematical form  \cite{sudheer2002data}.  Besides, generally speaking, the ANN  can handle  more complex nonlinear relationships than the traditional regression approaches  \cite{livingstone2008artificial},  so that the ANN is applicable across a wider range of problems.  Especially, the ANN can be easily applied to deal with  classification problems  with complicated boundary   \cite{bala2017classification}.  Therefore, we use the ANN here to model the relationship between the parameters of three-body problem and to classify the types of orbits and their stability.
	
	 	  In this paper,  we propose an effective approach  and  roadmap  to numerically gain  planar periodic orbits of three-body systems with arbitrary masses by means of machine learning based on an artificial neural network (ANN) model.  Given any a known periodic orbit as a starting point,  this approach can provide more and more periodic orbits (of the same family name) with  variable masses, while the mass domain having periodic orbits becomes larger and larger,  and the ANN model becomes wiser and wiser.   Finally we have an ANN model trained by means of all obtained periodic orbits of the same family, which provides a convenient way to give accurate enough periodic orbits with {\em arbitrary} masses for  physicists and astronomers.  In addition,  the periodicity and stability of orbits for {\em arbitrary} masses can be well predicted by an ANN classifier model.

	It must be emphasized that high performance computer and artificial intelligence (including machine learning) play important roles in solving periodic orbits of triple systems.   Today, nothing can prevent us from obtaining massive periodic solutions of three-body problem.  This is due to the great contributions of some great mathematicians, scientists and engineers in more than three hundred years!  Especially, it is Poincar\'{e}  \cite{Poincare1890} who made a historical turning point by proving the non-existence of the uniform first integral of triple system, which implies that we had to use numerical approach in general.  It is Turing  \cite{Turing1936, Turing1950} who published  his  epoch-making  papers  that  became  the foundation of modern computer and artificial intelligence.   It is  Von Neumann  \cite{VonNeumann1958} who proposed the so-called ``Von Neumann - Machine'' for modern computer.  It is  Jack S. Kilby, the winner of   Nobel prize for physics in 2000,  who took part in the invention of the integrated circuit.    And so on.   The famous three-body  problem  might be an excellent example to illustrate the importance of inventing new tools for human being to better understand and explore the nature.

	The approach and roadmap mentioned in this paper has general meanings.   Note that thousands of families of periodic orbits of triple systems with three or two equal masses have currently been found (for example, please visit the website \url{https://github.com/sjtu-liao/three-body}).   Using any of them as a starting point, we can similarly gain a trained ANN model to give accurate enough predictions of periodic orbits of the same family of triple system in a corresponding mass domain.  All of these might form a massive data base for periodic orbits of triple systems, which should be helpful to enrich and deepen our understandings about the famous three-body systems.   Besides, hopefully, some physical laws, such as generalized Kepler's law  \cite{Li2017}  for periodic orbits of triple systems with  {\em arbitrary} masses, could be found in future by analysing these massive data by means of machine learning.   
	
	Note also that nearly all known periodic orbits of triple systems  are planar, i.e. two-dimensional.  In theory,  the same ideas mentioned in this paper can be used to search for periodic orbits of triple systems in three dimensions.  Although, according to our results reported here and in the previous papers  \cite{Li2018, Li2020}, given a pair $(m_1,m_2)$, there is only one possible set $(x_1,v_1,v_2,T)$ that would lead to a periodic orbit of the three-body problem,   it should be interesting to investigate whether there exist bifurcations of periodic orbits in future.

\section*{Acknowledgements} Thanks a lot to the anonymous reviewers for their valuable comments.  This work was carried out on TH-1A (in Tianjin) at National Supercomputer Center, China.  It is partly supported by National Natural Science Foundation of China (Approval Nos. 91752104 and 12002132) and Shanghai Pilot Program for Basic Research - Shanghai Jiao Tong University ( No. 21TQ1400202).
 
\section*{Data Availability Statements} 
The data and the code related to this paper  can be seen in the website (\url{https://github.com/sjtu-liao/three-body}), 
including the trained ANN models of the two cases, all the periodic orbits found (also including those reported in   \cite{Li2017, Li2018, Li2020}), the code to train the ANN and use the ANN to predict the periodic orbits, and the code to train the ANN for classifying the orbits.
 
\section*{Author contributions} S.J.L. provided the main ideas and wrote the manuscript.  X.M.L. calculated the periodic orbits.   Y.Y.  analysed the results by means of machine learning based on an ANN model.  All authors contributed to the discussion and revision of the final manuscript.
 
\section*{Declaration of competing interest} The authors declare that they have no competing financial interests.
 

\section*{Appendix -- Direct search for periodic orbits with arbitrary masses}

The phase space of the planar three-body problem has 12 dimensions.
The grid searching method suffers the curse of dimensionality in such high dimensions. In order to reduce the dimensionality of the searching space, we can fix some parameters of initial conditions.
Since the collinear instant of three bodies is common in the three-body system, it is reasonable to consider three bodies are in a line at the start of periodic orbits.
The three bodies are in a line for the BHH family of periodic orbits:
$\bm{r}_1(0)=(x_1,0)$,  $\bm{r}_2(0)=(x_2,0)$,   $\bm{r}_3(0)=(x_3,0)$,
and their initial velocities are orthogonal to the  line:
 $\dot{\bm{r}}_1(0)=(0,v_1)$,  $\dot{\bm{r}}_2(0)=(0,v_2)$, $\dot{\bm{r}}_3(0)=(0, v_3)$.

According to the scaling law of the three-body system  \cite{Li2020}, we can fix $x_2=1$ and $x_3=0$. Without loss of generality, we assume momentum of the system is equal to zero, thus $v_3=-(m_1v_1+m_2v_2)/m_3$.

The approximated initial conditions and periods of periodic orbits can be  obtained through grid searching method, then these  approximated periodic orbits will be corrected by means of the Newton-Raphson method   \cite{Farantos1995, Lara2002,  Abad2011} and the clean numerical simulation (CNS)   \cite{Liao2009,  Liao2014-SciChina,  Hu2020}. The differential equations of the planar three-body problem can be described as:
\begin{equation}\label{odes}
        \dot{\bm{x}}=\bm{f}(\bm{x}); \quad \bm{x}(0)=\bm{y}, \quad \bm{x} \in \mathbb{R}^{12},
\end{equation}
and $\bm{x}=\bm{x}(t, \bm{y})$, $t \in \mathbb{R}$,  $\bm{x, y} \in \mathbb{R}^{12}$ is the solution of these equations, where $\bm{y}$ is the initial condition.

For the initial condition $\bm{y}$ and period $T$, a relative periodic orbit with a rotation angle $\theta$ means that 
\begin{equation}\label{def:periodic}
    \bm{x}(T, \bm{y})-\bm{P}(\theta)\bm{y}=0,
\end{equation}
where $\bm{P}(\theta)$ is the rotation matrix.

Now we correct the approximated initial conditions and periods of periodic orbits.
Let us assume the approximated initial conditions and periods are $(\bm{y}_i, T_i)$ at $i$-th step.
To improve accuracy of the approximated initial conditions and periods, the corrections of initial conditions and periods $(\Delta \bm{y}_i,\Delta T_i)$ should be satisfied the following equations:. 
\begin{equation}
\bm{x}(T_i+\Delta T_i, \bm{y}_i+\Delta \bm{y}_i)-\bm{P}(\theta)(\bm{y}_i+\Delta \bm{y}_i)=0.
\end{equation}
Using the first order Taylor approximation of the above equations, we have 
\begin{equation}
\bm{x}(T_i, \bm{y}_i)-\bm{P}(\theta)\bm{y}_i + \left(\frac{\partial \bm{x}}{\partial \bm{y}}-\bm{P}(\theta)\right)\Delta \bm{y}_i + \frac{\partial \bm{x}}{\partial t}\Delta T_i=0,
\end{equation}
where $\bm{P}(\theta)$ is the rotation matrix, $\bm{M}=\partial \bm{x}/\partial \bm{y}$ is the solution of variational equations. $\partial \bm{x}/\partial t$ is the derivative of the solution at $t=T_i$, i.e., $\bm{f}(\bm{y}_{T_i})$, where $\bm{y}_{T_i}=\bm{x}(T_i, \bm{y}_i)$.

Finally, we obtain the following linear system:
\begin{equation}\label{linearEquation}
    \left (
        \begin{array}{cc}
             \bm{M}-\bm{P}(\theta)     &   \bm{f}(\bm{y}_{T_i}) \\
        \end{array}
    \right )
    \left (
        \begin{array}{c}
        \Delta \bm{y}_i \\
        \Delta T_i
        \end{array}
    \right )=
    \left (
        \begin{array}{c}
            \bm{P}(\theta)\bm{y}_i-\bm{y}_{T_i} \\
        \end{array}
    \right ).
\end{equation}
Then the correction of initial conditions and periods of periodic orbits can be computed through solving these linear equations. 
As we fix some parameters of the initial conditions, only variables $x_1$, $v_1$, $v_2$ and $T$ will be modified. 
Then we have linear equations $\bm{A}\bm{z}=\bm{b}$, where $\bm{A}$ is a $12 \times 4$ matrix, $\bm{z}=(\Delta x_1\;\Delta v_1\; \Delta v_2 \;  \Delta T)^\intercal$ is a 4-vector and $\bm{b}$ is a 12-vector. Since the matrix $\bm{A}$ is not a square matrix, we solve this system by means of the least-norm method with SVD   \cite{Trefethen1997}.

\bibliography{ref-3body} 

\begin{thebibliography}{10}
\expandafter\ifx\csname url\endcsname\relax
  \def\url#1{\texttt{#1}}\fi
\expandafter\ifx\csname urlprefix\endcsname\relax\def\urlprefix{URL }\fi
\expandafter\ifx\csname href\endcsname\relax
  \def\href#1#2{#2} \def\path#1{#1}\fi

\bibitem{Newton1687}
I.~Newton, Philosophi{\ae} naturalis principia mathematica ({Mathematical
  Principles of Natural Philosophy}), London: Royal Society Press, 1687.

\bibitem{Poincare1890}
J.~H. Poincar{\' e}, {Sur} le probl{\' e}me des trois corps et les {\'
  e}quations de la dynamique. {D}ivergence des s{\' e}ries de m. {Lindstedt},
  Acta Math. 13 (1890) 1--270.

\bibitem{Broucke1975a}
R.~Broucke, On relative periodic solutions of the planar general three-body
  problem, Celestial Mechanics 12~(4) (1975) 439 -- 462.

\bibitem{Broucke1975b}
R.~Broucke, D.~Boggs, Periodic orbits in the planar general three-body problem,
  Celestial mechanics 11~(1) (1975) 13--38.

\bibitem{Hadjidemetriou1975a}
J.~D. Hadjidemetriou, The stability of periodic orbits in the three-body
  problem, Celestial Mechanics 12~(3) (1975) 255--276.

\bibitem{Henon1976}
M.~H{\'e}non, A family of periodic solutions of the planar three-body problem,
  and their stability, Celestial mechanics 13~(3) (1976) 267--285.

\bibitem{More1993}
C.~Moore, Braids in classical dynamics, Phys. Rev. Lett. 70 (1993) 3675--3679.

\bibitem{Montgomery1998}
R.~Montgomery, The {N}-body problem, the braid group, and action-minimizing
  periodic solutions, Nonlinearity 11~(2) (1998) 363.

\bibitem{Li2017}
X.~Li, S.~Liao, More than six hundred new families of {N}ewtonian periodic
  planar collisionless three-body orbits, Science China - Physics, Mechanics \&
  Astronomy 60~(12) (2017) 129511.

\bibitem{Li2018}
X.~Li, Y.~Jing, S.~Liao, Over a thousand new periodic orbits of a planar
  three-body system with unequal masses, Publications of the Astronomical
  Society of Japan 70~(4) (2018) 64.

\bibitem{Sun2018kepler}
B.~Sun, Kepler’s third law of n-body periodic orbits in a newtonian
  gravitation field, Science China Physics, Mechanics \& Astronomy 61~(5)
  (2018) 1--4.

\bibitem{Kwiecinski2018IJBC}
J.~A. Kwiecinski, A.~Kovacs, A.~L. Krause, F.~B. Planella, R.~A. Van~Gorder,
  Chaotic dynamics in the planar gravitational many-body problem with rigid
  body rotations, International Journal of Bifurcation and Chaos 28 (2018)
  1830013.

\bibitem{Stone2019}
N.~C. Stone, N.~W. Leigh, A statistical solution to the chaotic,
  non-hierarchical three-body problem, Nature 576~(7787) (2019) 406--410.

\bibitem{Li2019NA}
X.~Li, S.~Liao, Collisionless periodic orbits in the free-fall three-body
  problem, New Astronomy 70 (2019) 22--26.

\bibitem{Lin2019PRD}
K.~Lin, X.~Zhao, C.~Zhang, T.~Liu, B.~Wang, S.~Zhang, X.~Zhang, W.~Zhao,
  T.~Zhu, , A.~Wang, Gravitational waveforms, polarizations,response functions,
  and energy losses of triple systems in einstein-aether theory, Physical
  Review D 99 (2019) 023010.

\bibitem{Tanikawa2019CMDA}
K.~Tanikawa, M.~M. Saito, S.~Mikkola, A search for triple collision orbits
  inside the domain of the free-fall three-body problem, Celestial Mechanics
  and Dynamical Astronomy 131 (2019) 24.

\bibitem{Abouelmagd2020NA}
E.~I. Abouelmagd, J.~L.~G. Guirao, A.~K. Pal, Periodic solution of the
  nonlinear {S}itnikov restricted three-body problem, New Astronomy 75 (2020)
  101319.

\bibitem{Li2020}
X.~Li, X.~Li, S.~Liao, One family of 13315 stable periodic orbits of
  non-hierarchical unequal-mass triple systems, Science China - Physics,
  Mechanics \& Astronomy 64~(1) (2021) 219511.

\bibitem{Lorenz1963}
E.~Lorenz, Deterministic non-periodic flow, Journal of the Atmospheric Sciences
  20~(2) (1963) 130--141.

\bibitem{Turing1936}
A.~Turing, On computable numbers, with an application to the {E}ntscheidungs
  problem, Proc. London Maths. Soc. 42 (1936) 230–265.

\bibitem{Turing1950}
A.~Turing, Computing machinery and intelligence, Mind 50 (1950) 433–460.

\bibitem{VonNeumann1958}
J.~Von~Neumann, The Computer and the Brain, Yale University Press, 1958.

\bibitem{Suvakov2013}
M.~\ifmmode~\check{S}\else \v{S}\fi{}uvakov, V.~Dmitra\ifmmode \check{s}\else
  \v{s}\fi{}inovi\ifmmode~\acute{c}\else \'{c}\fi{}, Three classes of
  {N}ewtonian three-body planar periodic orbits, Phys. Rev. Lett. 110 (2013)
  114301.

\bibitem{Lorenz1989}
E.~Lorenz, Computational chaos-a prelude to computational instability, Physica
  D: Nonlinear Phenomena 35~(3) (1989) 299--317.

\bibitem{Lorenz2006}
E.~Lorenz, Computational periodicity as observed in a simple system, Tellus A
  58 (2006) 549 -- 557.

\bibitem{jianping2000}
J.~Li, Q.~Peng, J.~Zhou, Computational uncertainty principle in nonlinear
  ordinary differential equations (i), Science in China 43 (2000) 449--460.

\bibitem{jianping2001}
J.~Li, Q.~Zeng, J.~Chou, Computational uncertainty principle in nonlinear
  ordinary differential equations, Science in China Series E: Technological
  Sciences 44~(1) (2001) 55--74.

\bibitem{Teixeira2005}
J.~Teixeira, C.~A. Reynolds, K.~Judd, Time step sensitivity of nonlinear
  atmospheric models: numerical convergence, truncation error growth, and
  ensemble design, Journal of the Atmospheric Sciences 64~(1) (2007) 175--189.

\bibitem{Yao2008}
L.~Yao, D.~Hughes, Comment on``computational periodicity as observed in a
  simple system'' by {Edward N. Lorenz} (2006), Tellus A 60 (2008) 803 -- 805.

\bibitem{Chandramoorthy2021JCP}
N.~Chandramoorthy, Q.~Wang, On the probability of finding nonphysical solutions
  through shadowing, J. Comput. Phys. 440 (2021) 110389.

\bibitem{Liao2009}
S.~Liao, On the reliability of computed chaotic solutions of non-linear
  differential equations, Tellus A 61~(4) (2009) 550--564.

\bibitem{Liao2013-3b}
S.~Liao, Physical limit of prediction for chaotic motion of three-body problem,
  Communications in Nonlinear Science and Numerical Simulation 19~(3) (2014)
  601--616.

\bibitem{Liao2014-SciChina}
S.~Liao, P.~Wang, On the mathematically reliable long-term simulation of
  chaotic solutions of {L}orenz equation in the interval [0,10000], Sci. China
  - Phys. Mech. Astron. 57 (2014) 330 -- 335,.

\bibitem{Liao2015-IJBC}
S.~Liao, X.~Li, On the inherent self-excited macroscopic randomness of chaotic
  three-body systems, Int. J. Bifurcation \& Chaos 25 (2015) 1530023 (11 page).

\bibitem{Hu2020}
T.~Hu, S.~Liao, On the risks of using double precision in numerical simulations
  of spatio-temporal chaos, Journal of Computational Physics 418 (2020) 109629.

\bibitem{Liao2022AAMM}
S.~Liao, S.~Qin, Ultra-chaos: an insurmountable objective obstacle of
  reproducibility and replication, Adv. Appl. Math. Mech. 14~(4) (2022)
  799--815.

\bibitem{Farantos1995}
S.~C. Farantos, Methods for locating periodic orbits in highly unstable
  systems, Journal of Molecular Structure: THEOCHEM 341~(1) (1995) 91 -- 100.

\bibitem{Lara2002}
M.~Lara, J.~Pelaez, On the numerical continuation of periodic orbits - an
  intrinsic, 3-dimensional, differential, predictor-corrector algorithm,
  Astronomy and Astrophysics 389~(2) (2002) 692--701.

\bibitem{Abad2011}
A.~Abad, R.~Barrio, A.~Dena, Computing periodic orbits with arbitrary
  precision, Phys. Rev. E 84 (2011) 016701.

\bibitem{Allgower2003}
E.~L. Allgower, K.~Georg, Introduction to numerical continuation methods,
  Vol.~45, SIAM, 2003.

\bibitem{Jankovic2016}
M.~R. Jankovi\ifmmode~\acute{c}\else \'{c}\fi{}, V.~Dmitra\ifmmode
  \check{s}\else \v{s}\fi{}inovi\ifmmode~\acute{c}\else \'{c}\fi{}, Angular
  momentum and topological dependence of {Kepler's} third law in the
  {B}roucke-{H}adjidemetriou-{H}\'enon family of periodic three-body orbits,
  Phys. Rev. Lett. 116 (2016) 064301.

\bibitem{Jankovic2020}
M.~R. Jankovi{\'c}, V.~Dmitra{\v{s}}inovi{\'c}, M.~{\v{S}}uvakov, A guide to
  hunting periodic three-body orbits with non-vanishing angular momentum,
  Computer Physics Communications 250 (2020) 107052.

\bibitem{hassoun1995fundamentals}
M.~H. Hassoun, et~al., Fundamentals of artificial neural networks, MIT press,
  1995.

\bibitem{gevrey2003review}
M.~Gevrey, I.~Dimopoulos, S.~Lek, Review and comparison of methods to study the
  contribution of variables in artificial neural network models, Ecological
  modelling 160~(3) (2003) 249--264.

\bibitem{andrews1995survey}
R.~Andrews, J.~Diederich, A.~B. Tickle, Survey and critique of techniques for
  extracting rules from trained artificial neural networks, Knowledge-based
  systems 8~(6) (1995) 373--389.

\bibitem{abiodun2018state}
O.~I. Abiodun, A.~Jantan, A.~E. Omolara, K.~V. Dada, N.~A. Mohamed, H.~Arshad,
  State-of-the-art in artificial neural network applications: A survey, Heliyon
  4~(11) (2018) e00938.

\bibitem{sudheer2002data}
K.~Sudheer, A.~Gosain, K.~Ramasastri, A data-driven algorithm for constructing
  artificial neural network rainfall-runoff models, Hydrological processes
  16~(6) (2002) 1325--1330.

\bibitem{livingstone2008artificial}
D.~J. Livingstone, Artificial neural networks: methods and applications,
  Springer, 2008.

\bibitem{bala2017classification}
R.~Bala, D.~Kumar, Classification using ann: A review, Int. J. Comput. Intell.
  Res 13~(7) (2017) 1811--1820.

\bibitem{reddi2019convergence}
S.~J. Reddi, S.~Kale, S.~Kumar, On the convergence of {Adam} and beyond, arXiv
  preprint arXiv:1904.09237.

\bibitem{Padmanabhan1985}
T.~Padmanabhan, Physical significance of {P}lanck length, Annals of Physics 165
  (1985) 38 -- 58.

\bibitem{kapela2007}
T.~Kapela, C.~Sim{\'o}, Computer assisted proofs for nonsymmetric planar
  choreographies and for stability of the {Eight}, Nonlinearity 20~(5) (2007)
  1241.

\bibitem{Simo2002}
C.~Sim{\'o}, Dynamical properties of the figure eight solution of the
  three-body problem, Celestial Mechanics: Dedicated to Donald Saari for His
  60th Birthday: Proceedings of an International Conference on Celestial
  Mechanics, December 15-19, 1999, Northwestern University, Evanston, Illinois
  292 (2002) 209.

\bibitem{prechelt1998early}
L.~Prechelt, Early stopping-but when?, in: Neural Networks: Tricks of the
  trade, Springer, 1998, pp. 55--69.

\bibitem{Dimitrov2017}
W.~Dimitrov, H.~Lehmann, K.~Kami{\'n}ski, M.~K. Kami{\'n}ska, M.~Zg{\'o}rz,
  M.~Gibowski, The hierarchical triple system {DY Lyncis}, Monthly Notices of
  the Royal Astronomical Society 466~(1) (2017) 2--10.

\bibitem{Trefethen1997}
L.~Trefethen, D.~Bau~III, Numerical Linear Algebra, Society for Industrial and
  Applied Mathematics, Philadelphia, PA, 1997.

\end{thebibliography}

\end{document}